\documentclass{swsc}
\usepackage{graphicx}
\usepackage{bm,natbib,url,wasysym}
\graphicspath{{./fig/}{./png/}}
\newcommand{\EQ}{\begin{equation}}
\newcommand{\EN}{\end{equation}}
\newcommand{\EQA}{\begin{eqnarray}}
\newcommand{\ENA}{\end{eqnarray}}

\newcommand{\Eq}[1]{Equation~(\ref{#1})}
\newcommand{\Eqs}[2]{Equations~(\ref{#1}) and~(\ref{#2})}

\newcommand{\Sec}[1]{Section~\ref{#1}}

\newcommand{\Fig}[1]{Figure~\ref{#1}}

\newcommand{\Tab}[1]{Table~\ref{#1}}

\newcommand{\bra}[1]{\langle #1\rangle}

{}
{}
{}

{}
{}
{}
{}
{}
{}
{}
{}
\newcommand{\meanAA}{\overline{\mbox{\boldmath $A$}}{}}{}
\newcommand{\meanBB}{\overline{\mbox{\boldmath $B$}}{}}{}
{}
\newcommand{\meanFF}{\overline{\mbox{\boldmath $F$}}{}}{}
\newcommand{\meanFFf}{\overline{\mbox{\boldmath $F$}}_{\rm f}{}}{}
\newcommand{\meanFFm}{\overline{\mbox{\boldmath $F$}}_{\rm m}{}}{}
{}
{}
{}
\newcommand{\meanJJ}{\overline{\mbox{\boldmath $J$}}{}}{}
\newcommand{\meanUU}{\overline{\bm{U}}}

{}
{}
{}

\newcommand{\meanh}{\overline{h}}
\newcommand{\meanhm}{\overline{h}_{\rm m}}
\newcommand{\meanhf}{\overline{h}_{\rm f}}

{}

{}
{}

\newcommand{\Rsun}{R}
\newcommand{\uu}{\mbox{\boldmath $u$} {}}
\newcommand{\UU}{\mbox{\boldmath $U$} {}}

\newcommand{\bb}{\mbox{\boldmath $b$} {}}
\newcommand{\BB}{\mbox{\boldmath $B$} {}}

\newcommand{\jj}{\mbox{\boldmath $j$} {}}
\newcommand{\JJ}{\mbox{\boldmath $J$} {}}

\newcommand{\AAA}{\mbox{\boldmath $A$} {}}
\newcommand{\aaaa}{\mbox{\boldmath $a$} {}}

\newcommand{\FF}{\mbox{\boldmath $F$} {}}

\newcommand{\nab}{\mbox{\boldmath $\nabla$} {}}

\newcommand{\dd}{{\rm d} {}}

\newcommand{\const}{{\rm const}  {}}

\def\kf{k_{\rm f}}

\def\urms{u_{\rm rms}}

\def\etat{\eta_{\rm t}}

\def\twothird{{\textstyle{2\over3}}}
\def\fourthird{{\textstyle{4\over3}}}

\newcommand{\AU}{\,{\rm AU}}

\newcommand{\yjgr}[3]{ #1, {J.\ Geophys.\ Res.,} {#2}, #3}
\newcommand{\ysol}[3]{ #1, {Sol.\ Phys.,} {#2}, #3}
\newcommand{\yapj}[3]{ #1, {ApJ,} {#2}, #3}

\newcommand{\yapjl}[3]{ #1, {ApJ,} {#2}, #3}

\newcommand{\yan}[3]{ #1, {Astron.\ Nachr.,} {#2}, #3}

\newcommand{\yana}[3]{ #1, {A\&A,} {#2}, #3}

\newcommand{\ygafd}[3]{ #1, {Geophys.\ Astrophys.\ Fluid Dyn.,} {#2}, #3}

\newcommand{\yjfm}[3]{ #1, {J.\ Fluid Mech.,} {#2}, #3}

\newcommand{\yprl}[3]{ #1, {Phys.\ Rev.\ Lett.,} {#2}, #3}

\newcommand{\ymn}[3]{ #1, {MNRAS,} {#2}, #3}

\newcommand{\ysph}[3]{ #1, {Sol. Phys.,} {#2}, #3}

\newcommand{\ybook}[3]{ #1, {#2} (#3)}

\newcommand{\sapjl}[1]{ #1, {ApJL}, submitted}

\begin{document}

\titlerunning{Magnetic twist: a source and property of space weather}
\authorrunning{J. Warnecke et al.}

\title{Magnetic twist: a source and property of space weather}
\author{J\"orn Warnecke\inst{1,2} \and Axel Brandenburg\inst{1,2} \and
  Dhrubaditya Mitra\inst{1}}
\institute{Nordita, AlbaNova University Center, Roslagstullsbacken 23,
SE-10691 Stockholm, Sweden
\and Department of Astronomy, AlbaNova University Center,
Stockholm University, SE-10691 Stockholm, Sweden}

\date{\today,~ $ $Revision: 1.62 $ $}
\abstract{}{%
We present evidence for finite magnetic helicity density in the heliosphere
and numerical models thereof, and relate it to the magnetic field properties
of the dynamo in the solar convection zone.
}{%
We use simulations and solar wind data to compute magnetic helicity either
directly from the simulations, or indirectly using time series of the
skew-symmetric components of the magnetic correlation tensor.
}{%
We find that the solar dynamo produces negative magnetic helicity at
small scales and positive at large scales.
However, in the heliosphere these properties are reversed and the
magnetic helicity is now positive at small scales and negative at large scales.
We explain this by the fact that a negative diffusive magnetic helicity
flux corresponds to a positive gradient of magnetic helicity, which leads
to a change of sign from negative to positive values at some radius
in the northern hemisphere.
}{}

\maketitle

\section{Introduction}

The magnetic field in the heliosphere is a direct consequence of the
solar dynamo converting kinetic energy of the convection zone into
magnetic energy.
The magnetic field is cyclic with a period of 22 years on average,
but has also significant fluctuations on top of this.
These fluctuation can be large enough to suppress the number of sunspots
to minimum levels for decades, for example during the Maunder minimum.
This minimum has been associated with the little ice age in the early 17th
century, although the solar activity to Earth climate relation remains
ill understood.
Of particular interest for space weather are strong variations
caused by coronal mass ejections (CMEs).
These events are believed to be a result of footpoint motions
of the magnetic field at the solar surface, driving strongly
stressed magnetic field configurations to a point when they become
unstable and release the resulting energy in an instant.
CMEs can shed large clouds of magnetized plasma into
interplanetary space and can accelerate charged particles to high
velocities towards the Earth.
The main driver of these ejections is the magnetic field, where the
energy of the eruption is stored.

Large-scale dynamos, for example the one operating in the solar
convection zone, produces magnetic helicity of opposite sign at
large and small scale.
Here, magnetic helicity is the dot product of the magnetic
field and the vector potential integrated over a certain volume.
For a long time it was believed that CMEs are disconnected from
the actual dynamo process, but this view has changed in the past 10 years.
In the regime of large magnetic Reynolds numbers, or high electric
conductivity, the magnetic helicity associated with the small-scale
field, quenches the dynamo \citep{PFL}.
This is a concept that is now well demonstrated using periodic
box simulations of helically forced turbulence \citep{B01}.
However, astrophysical large-scale dynamos are inhomogeneous
and drive magnetic helicity fluxes, whose divergence is
relevant for alleviating what is now often referred to as
catastrophic quenching of the large-scale dynamo.
An important fraction of these magnetic helicity fluxes is
associated with motions through the solar surface and their eventual
ejection into the interplanetary space \citep{BB03}.
Connecting the dynamo with the physics at and above the
solar surface is therefore an essential piece of dynamo physics.

In this paper we review the state of such models and their
ability to shed magnetic helicity and to produce ejections
of the type seen in the Sun.
We begin by discussing a simpler model in Cartesian geometry
and turn then to models in spherical wedges.
Finally, we compare with observations of magnetic helicity in
the solar wind and discuss our finding in connection with earlier
dynamo models.
Full details of this work have been published elsewhere
\citep{WB10,WBM11,BSBG11}, but here we focus on an aspect
that is common to all these papers, namely the nature of
magnetic twist associated with the ejecta away from the Sun.

\section{Plasmoid ejections in Cartesian models}
\label{Plasmoid}

A straightforward extension of dynamos in Cartesian domains
is to add an extra layer on top of it that mimics a nearly
force-free solar corona above it.
This was done by \cite{WB10} who used a dynamo that was driven
by turbulence that in turn was driven by a forcing function
in the momentum equation.
To imitate the effects of stratification and rotation that are
known to produce helicity, they used a forcing function
that was itself helical.
This leads to large-scale dynamos that are more efficient
than naturally occurring ones that are driven, for example,
by rotating convection \citep[see, e.g.,][]{KKBMT10,WKMB12,KMB12}.
In those papers, the produced kinetic helicity is much weaker.
As mentioned in the beginning, such dynamos can produce
magnetic fields whose small-scale contribution has
magnetic helicity of the same sign as that of the forcing function
and whose large-scale contribution has
magnetic helicity of opposite sign.

In our Cartesian model, the two horizontal directions ($x$ and $y$)
are equivalent, but when the large-scale field saturates, it must finally
choose one of the two possible directions.
This is a matter of chance but, in the case discussed below, the field
shows a large-scale variation in the $y$ direction.
The large-scale field settles into a state with minimal horizontal
wavenumber, which is here $(k_x,k_y)=(0,k_1)$, where $k_1=2\pi/L$ and
$L$ is the horizontal extent of the domain.
Note that for fields with variation along the diagonal the wavenumber
would be $\sqrt{2}$ times larger, so such a state is less preferred.

\begin{figure}[t!]\begin{center}
\includegraphics[width=.8\columnwidth, angle=92]{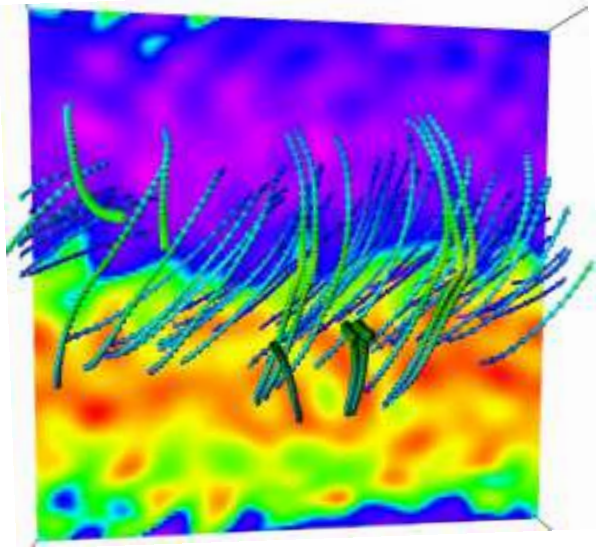}
\end{center}\caption[]{
Three-dimensional visualization of the magnetic field viewed from above.
The vertical magnetic field component is color-coded (light/yellow pointing
upward and dark/blue pointing downward).
Note that the field lines form a left-handed spiral over the scale of
the domain, as expected for turbulence with positive helicity at
small scales.
The $x$-axis points to the right while the $y$-axis points upward.
Adapted from \cite{WB10}.
}\label{twist}
\end{figure}

In \Fig{twist} we show the surface magnetic field of such a
dynamo of the work of \cite{WB10}.
We show color-coded the vertical (line of sight) magnetic
field component together with a perspective view of field lines
in the volume above, which we shall refer to as the corona region.
In addition to fluctuations, we can see a large-scale pattern
of the field with a sinusoidal modulation in the $x$ direction
and no systematic variation in the $y$ direction.
The field lines in the corona region show a spiraling pattern
corresponding to a left-handed spiral.
This is because the helicity of the forcing function is in this model
positive (right handed), so the resulting large-scale field must have
helicity of the opposite sign.

\begin{figure}[t!]\begin{center}
\includegraphics[width=.8\columnwidth]{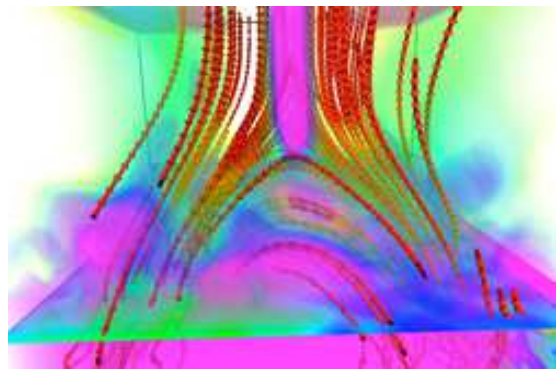}
\end{center}\caption[]{
Magnetic field structure in the dynamo exterior.
Field lines are shown in red and the modulus of the current density
is shown in pink with semi-transparent opacity.
Note the formation of a vertical current sheet above the arcade.
Adapted from \cite{WB10}.
}\label{fig_struc}
\end{figure}

\begin{figure}[t!]\begin{center}
\includegraphics[width=\columnwidth]{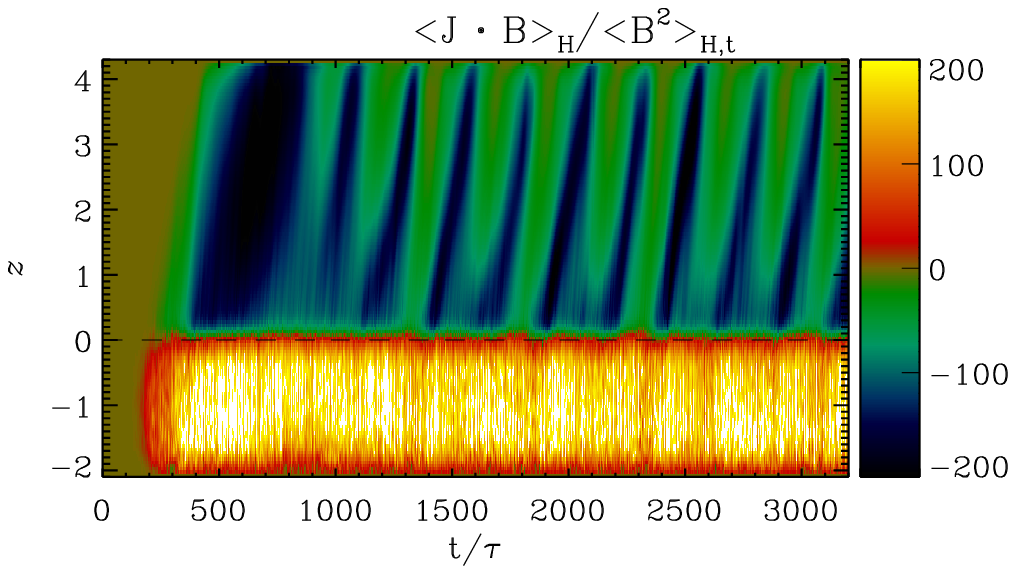}
\end{center}\caption[]{
Dependence of $\bra{\JJ\cdot\BB}_{\rm H}/\bra{\BB^2}_{\rm H,t}$
on time and height.
Dark/blue stands for negative and light/yellow for positive values.
For this run the vertical extent of the domain is $-\twothird\pi\leq
z\leq \fourthird\pi$.
}
\label{pjbm_cont_TT_ct5}
\end{figure}

\begin{figure*}[t!]\begin{center}
\includegraphics[width=0.5\columnwidth]{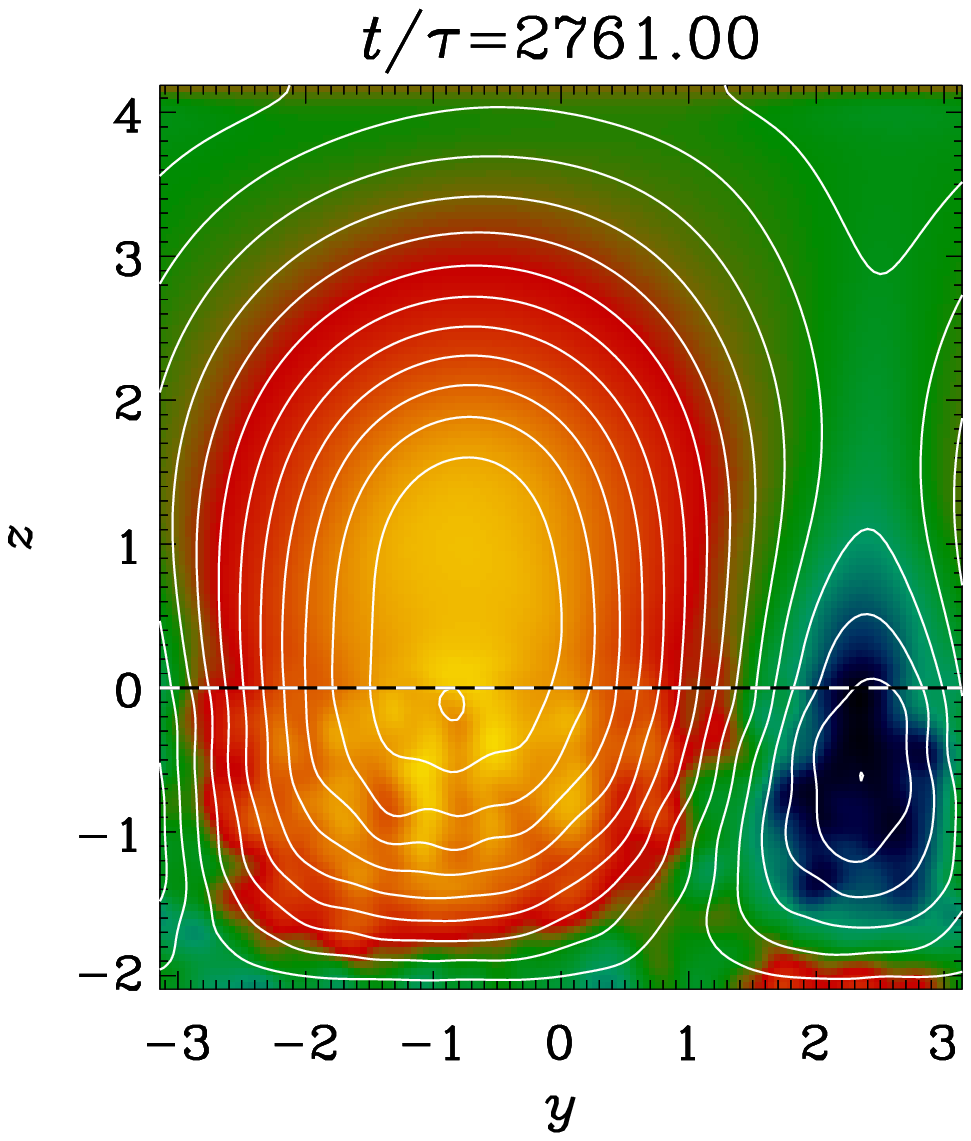}
\includegraphics[width=0.5\columnwidth]{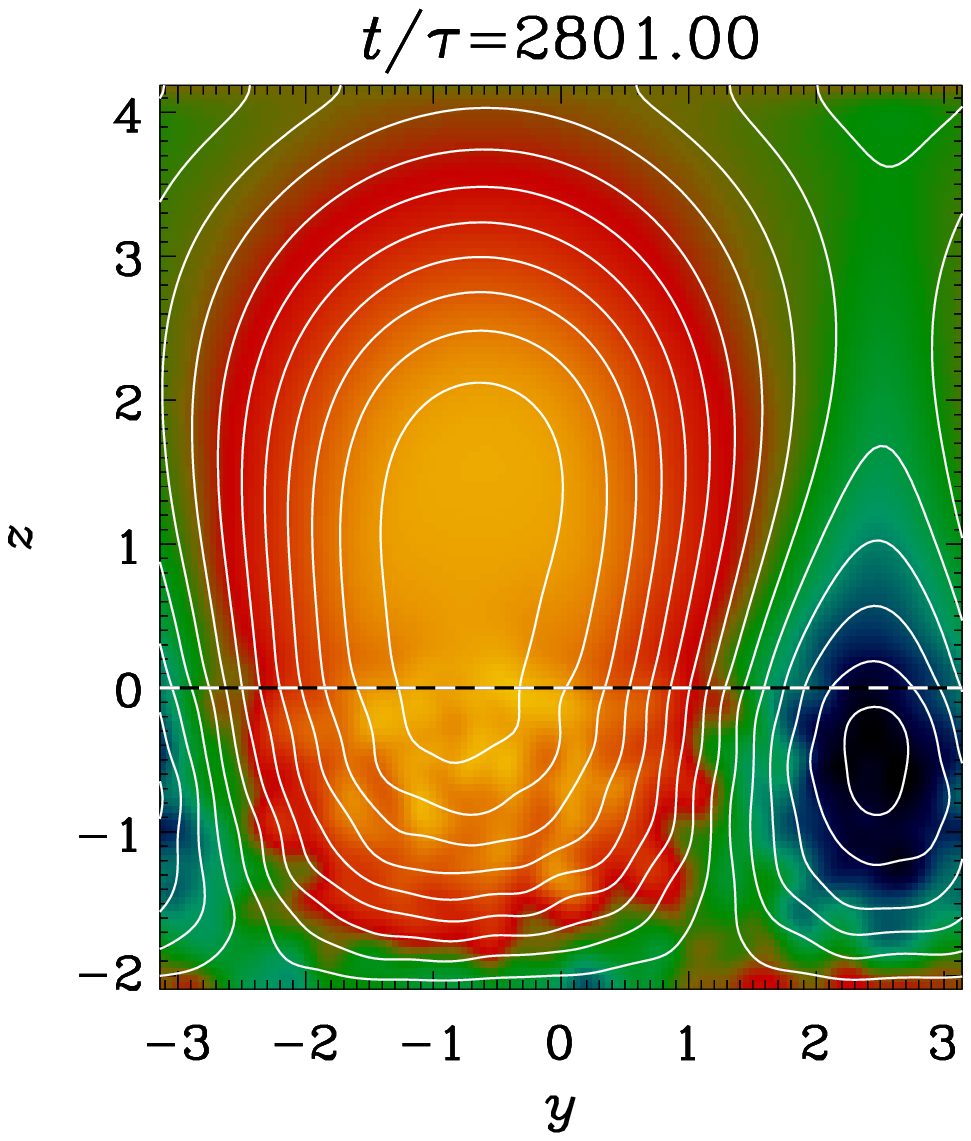}
\includegraphics[width=0.5\columnwidth]{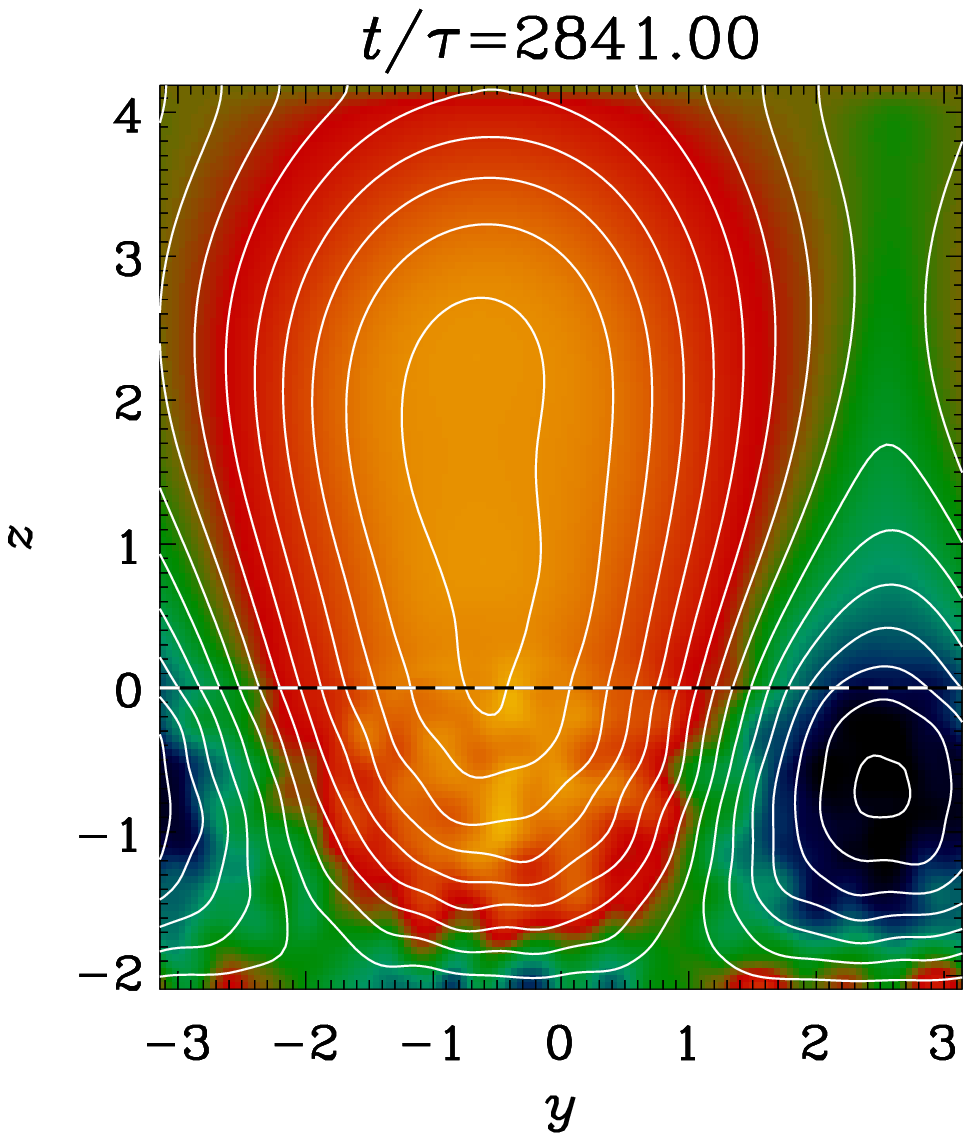}
\includegraphics[width=0.5\columnwidth]{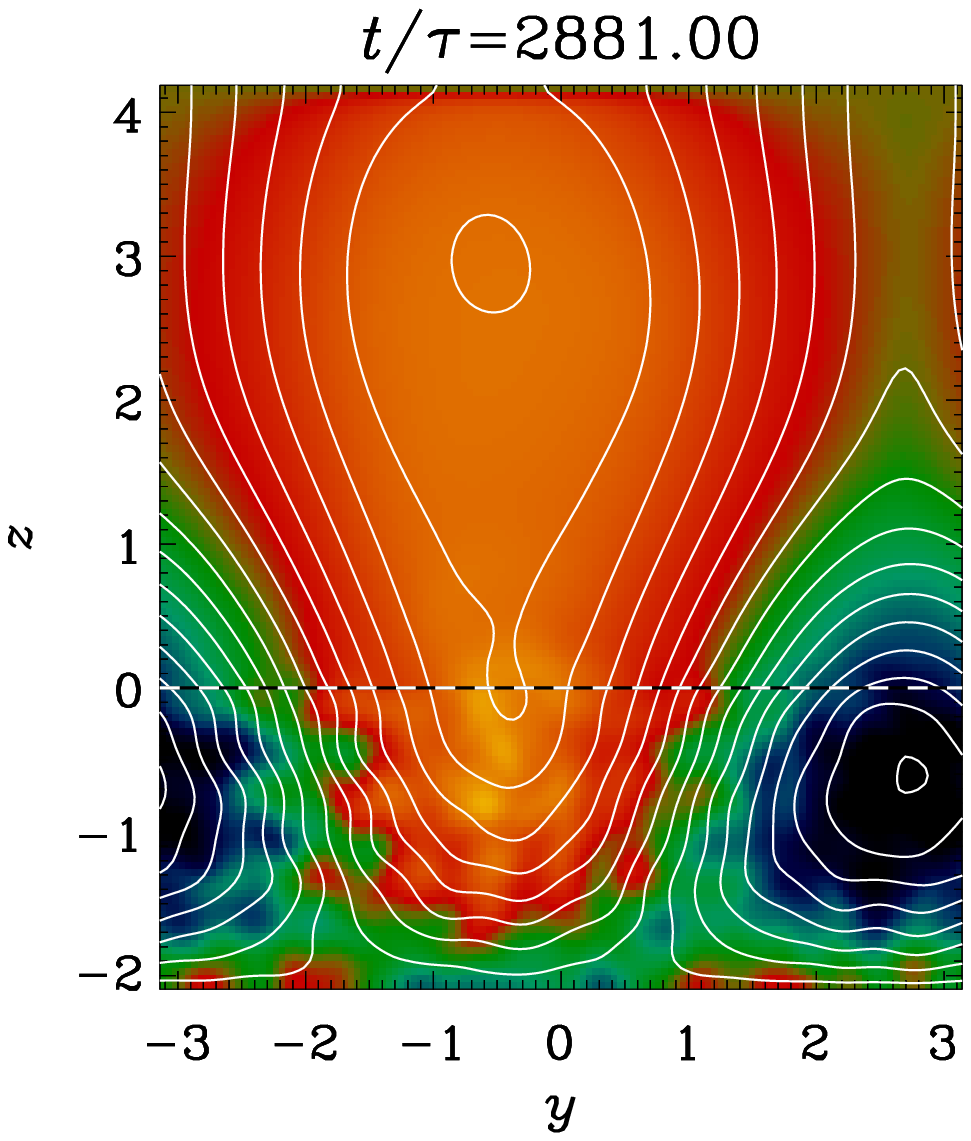}
\includegraphics[width=0.5\columnwidth]{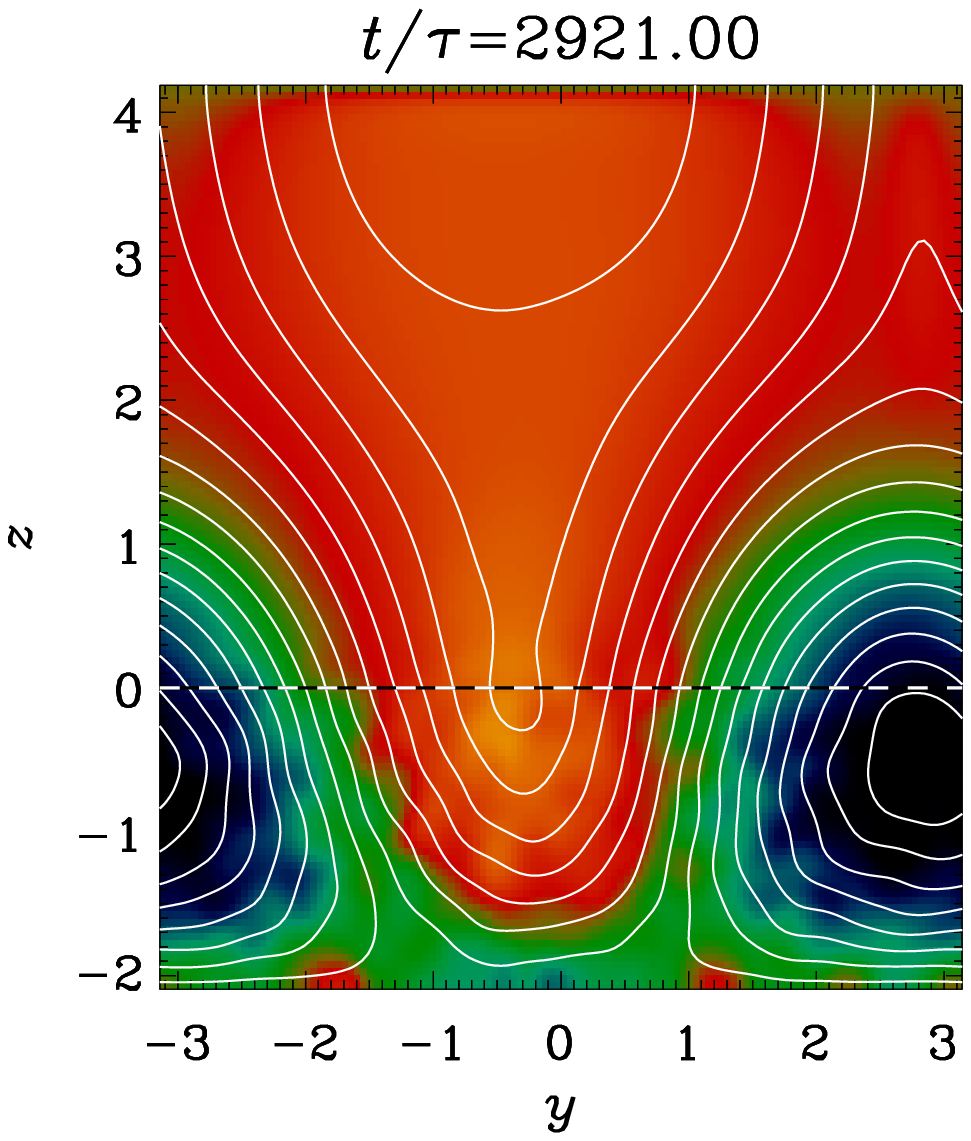}
\includegraphics[width=0.5\columnwidth]{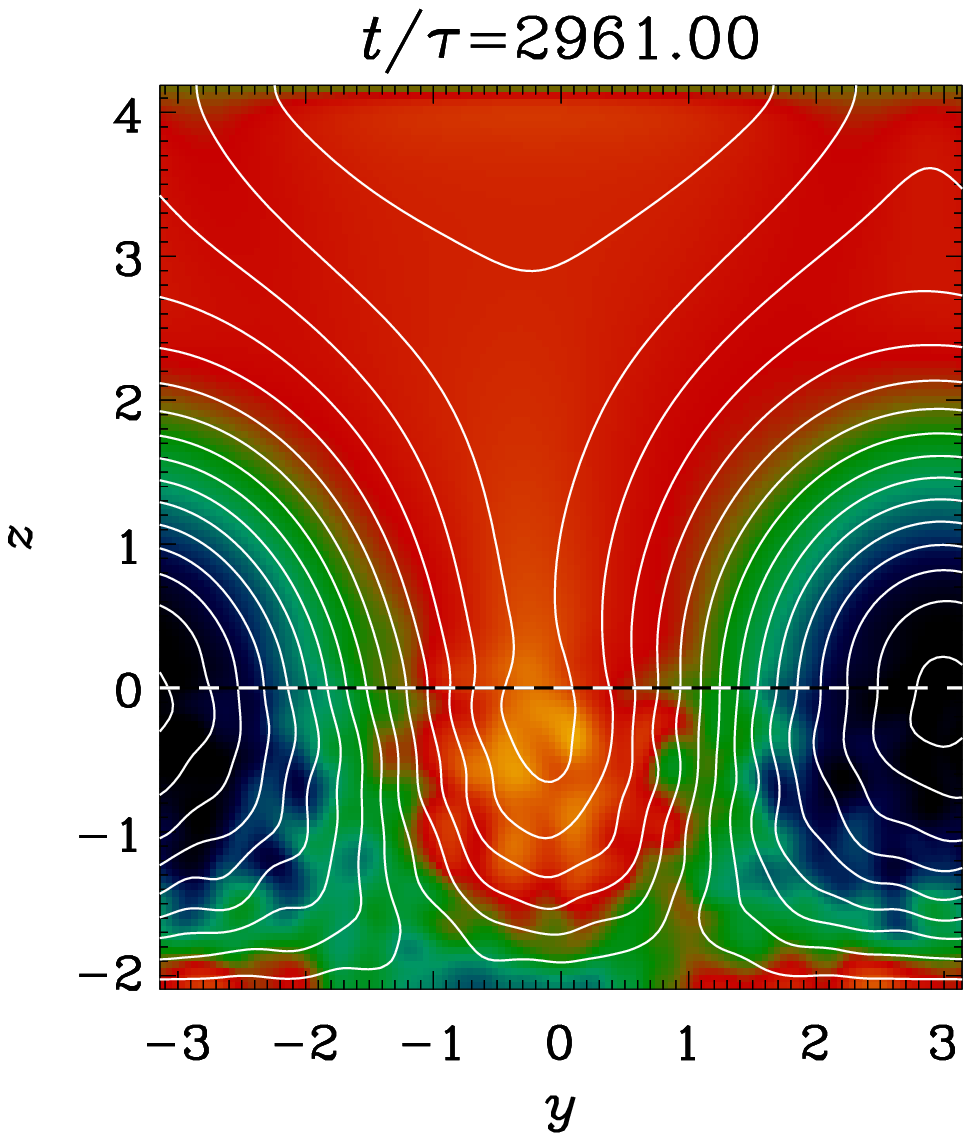}
\includegraphics[width=0.5\columnwidth]{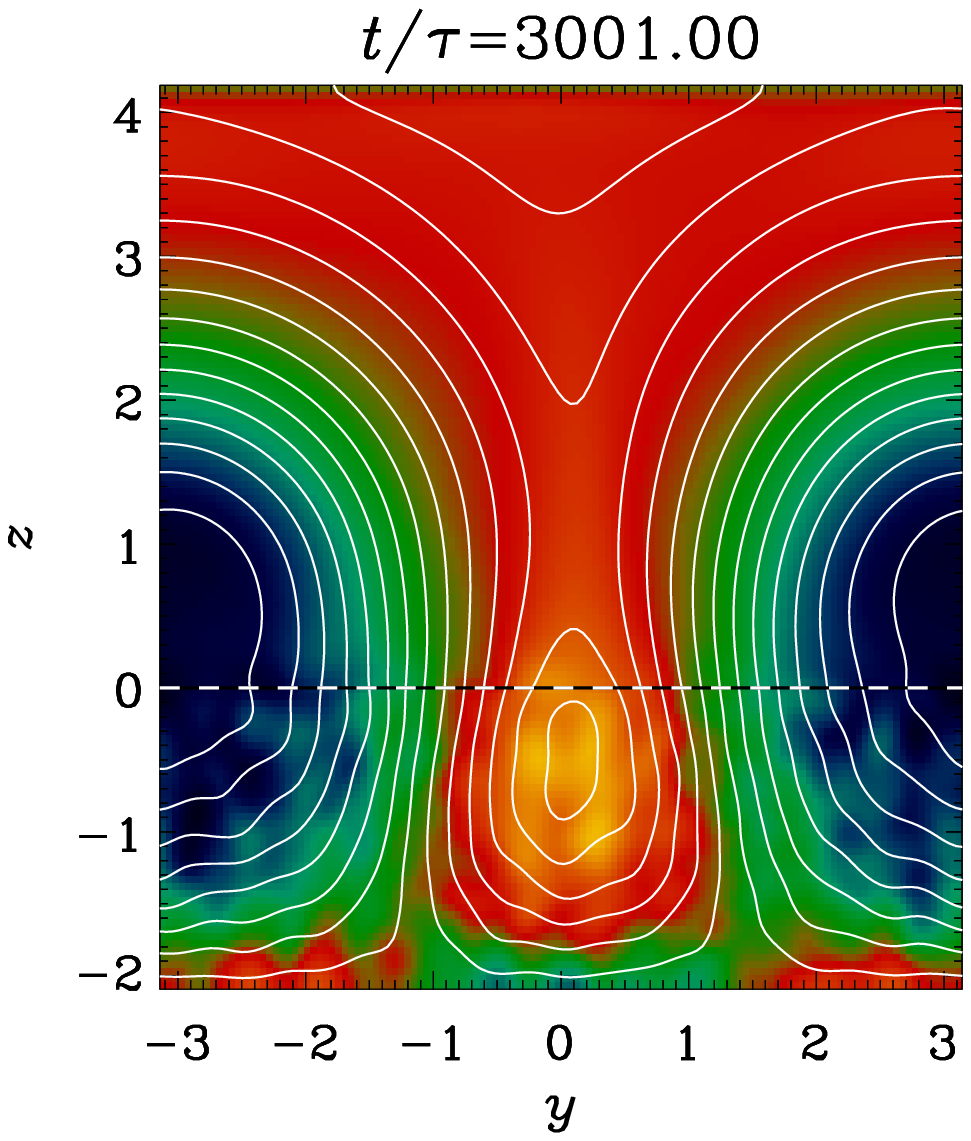}
\includegraphics[width=0.5\columnwidth]{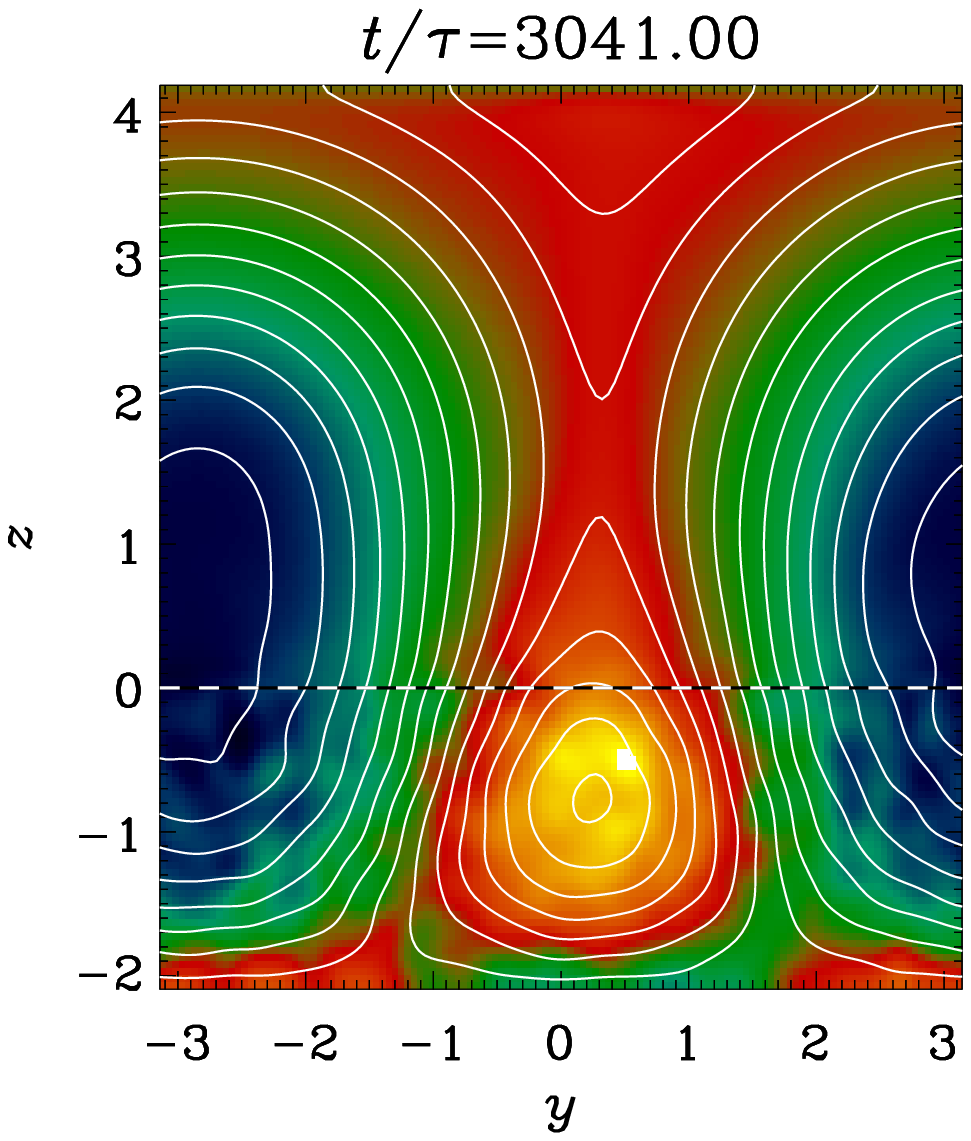}
\end{center}\caption[]{
Time series of the formation of a plasmoid ejection.
Contours of $\bra{A_x}_{x}$ are shown together with a
color-scale representation of $\bra{B_x}_x$;
dark/blue stands for negative and light/yellow for positive values,
as in \Fig{pjbm_cont_TT_ct5}.
The contours of $\bra{A_x}_{x}$ correspond to field lines of $\bra{\BB}_x$
in the $yz$ plane.
The dashed horizontal lines show the location of the surface at $z=0$.
For this run the vertical extent of the domain is $-\twothird\pi\leq z\leq {4\over3}\pi$.
}
\label{aax2_1}
\end{figure*}

The large-scale magnetic field is essentially steady in the dynamo
region and does not change its overall bipolar structure;
see \Fig{fig_struc} for a perspective view.
However the magnetic field of the corona region shows a time-dependent
oscillating structure associated with nearly regularly occurring ejection.
The ejection events can be monitored in terms of the current helicity,
$\JJ\cdot\BB$, where $\JJ=\nab\times\BB/\mu_0$
is the current density and $\mu_0$ the vacuum permeability.
To compensate for the radial decline of $\JJ\cdot\BB$, we have scaled
it by $\bra{\BB^2}_{H,t}$, which is here understood as a combined
average over horizontal directions and over time.
The result is shown in \Fig{pjbm_cont_TT_ct5}.

It is remarkable that the field does not remain steady in the outer parts.
This can be seen more clearly in a sequence of field line visualizations
in \Fig{aax2_1}.
Here, the magnetic field is averaged over the $x$-direction and we show $\bra{B_x}_{x}$
color coded together with magnetic field lines as contours of
$\bra{A_x}_{x}$ in the $yz$-plane.
Light/yellow shades correspond to positive values, the dark/blue to
negative, similar to \Fig{pjbm_cont_TT_ct5}.
Note that a concentration in $B_x$ emerges from the lower region to the outer one.
The magnetic field line surround the concentration and form a shape
similar to plasmoid ejections, which are believed to be a two-dimensional
model of producing CMEs \citep{OS93}.
At a time of $t/\tau=2881$ turnover times, the concentration is
split into two parts, where
the upper one leaves the domain through the upper boundary, while the
lower one stays in the lower layer.
Here, $\tau=(\urms\kf)^{-1}$ is the turnover time
based on the forcing wavenumber $\kf$ and
$\urms$ is the root mean squared velocity averaged in the lower layer.
At $t/\tau=3041$, the field line have formed an {\sf X}-point in the
center of the upper layer.
In an {\sf X}-point, field lines reconnect and release large amounts of
energy through Ohmic heating.
In the Sun these reconnection events are believed to trigger an
eruptive flare, which can cause a CME.
A similar behavior can also be seen in more realistic models in
spherical geometry, as will be discussed in the next section.

\section{Helicity reversals in spherical models}

\begin{figure*}[t!]\begin{center}
\includegraphics[width=3.3cm]{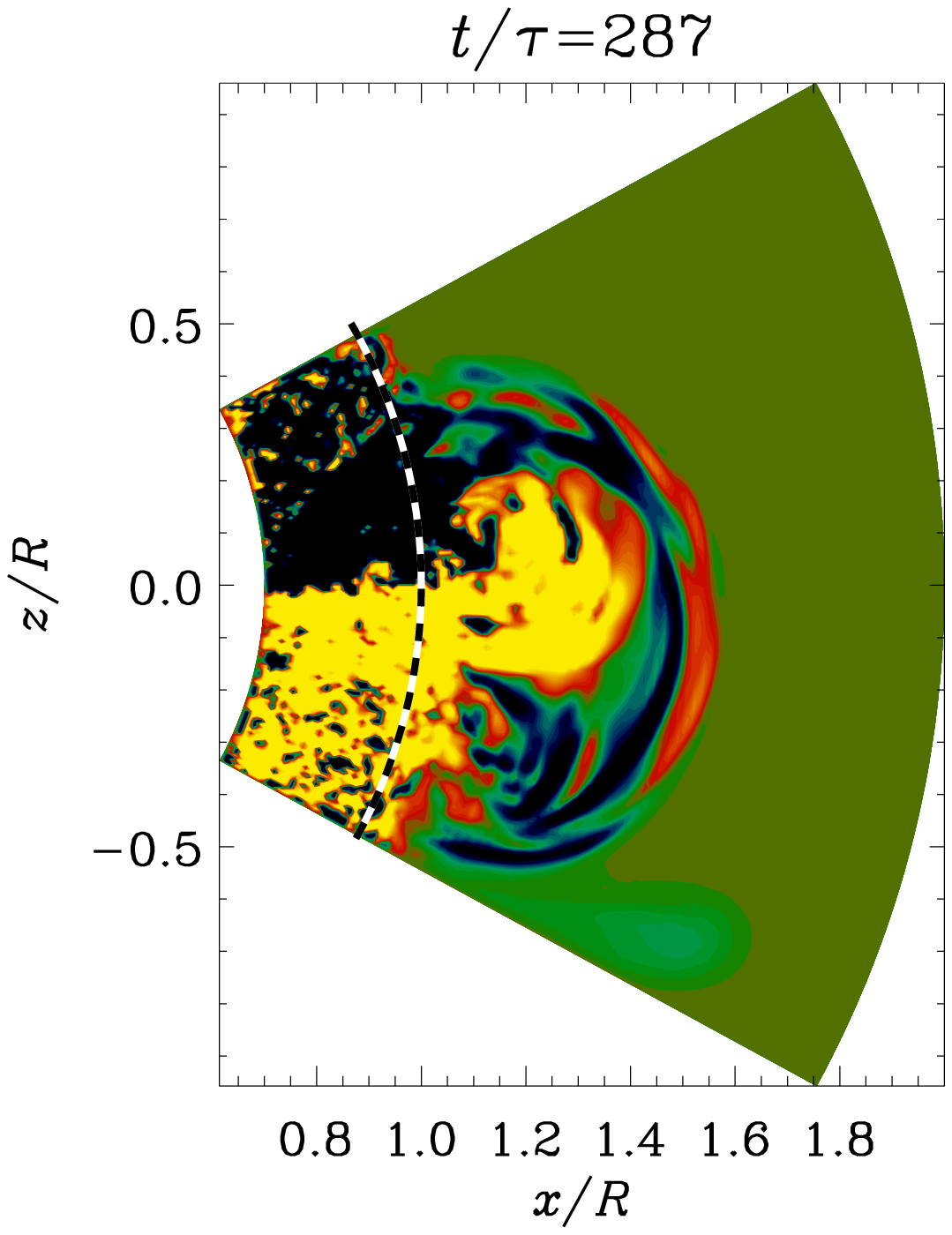}
\includegraphics[width=3.3cm]{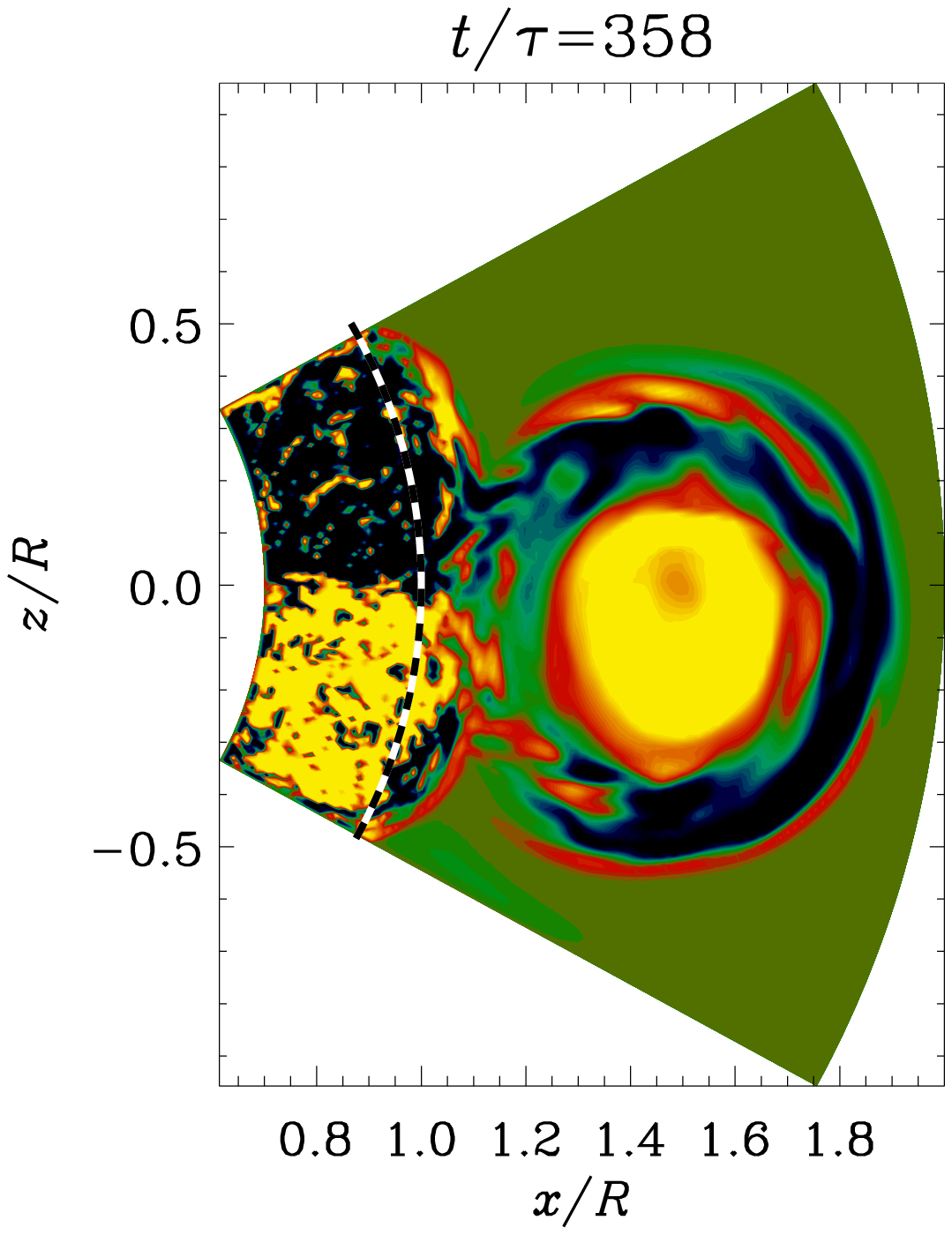}
\includegraphics[width=3.3cm]{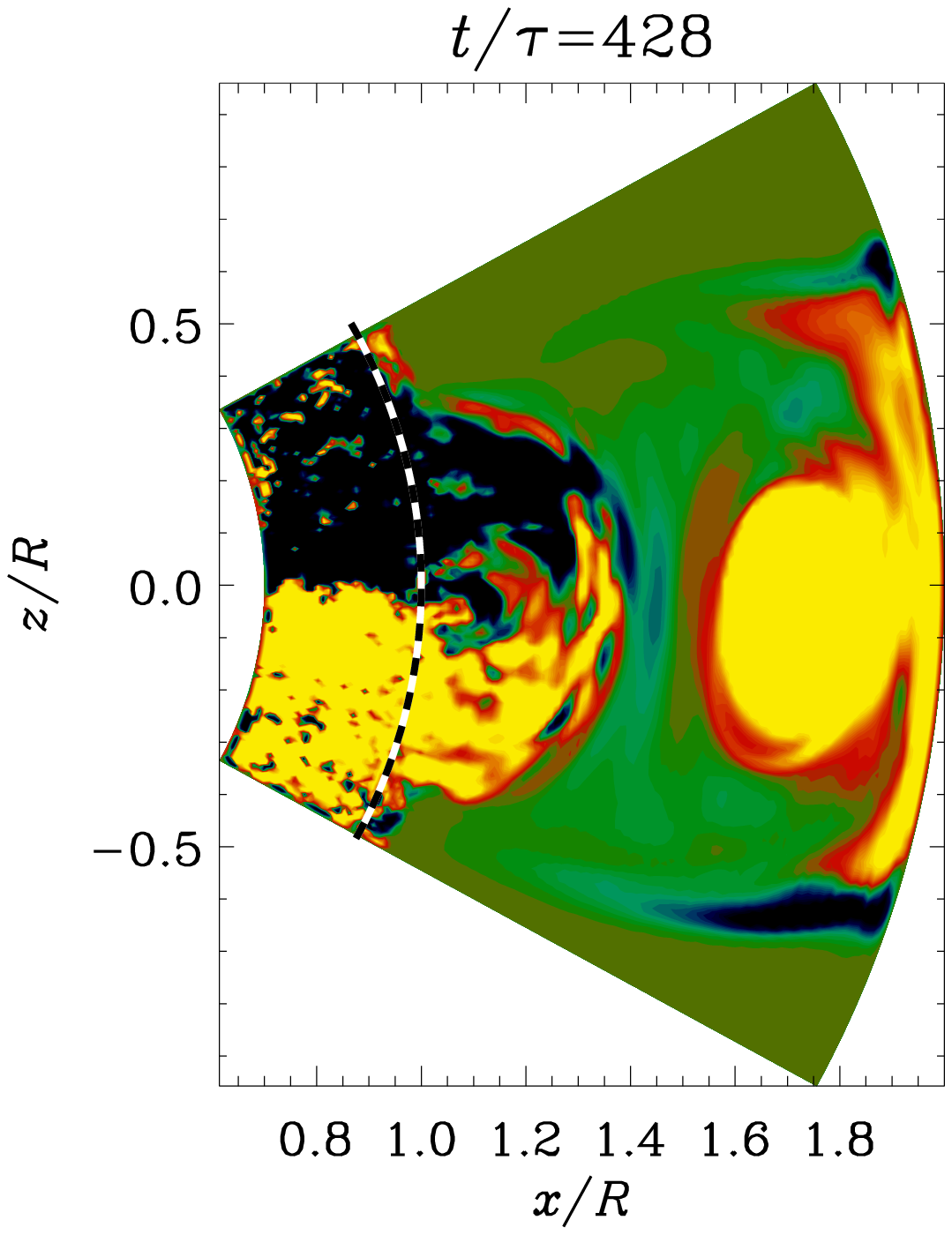}
\includegraphics[width=3.3cm]{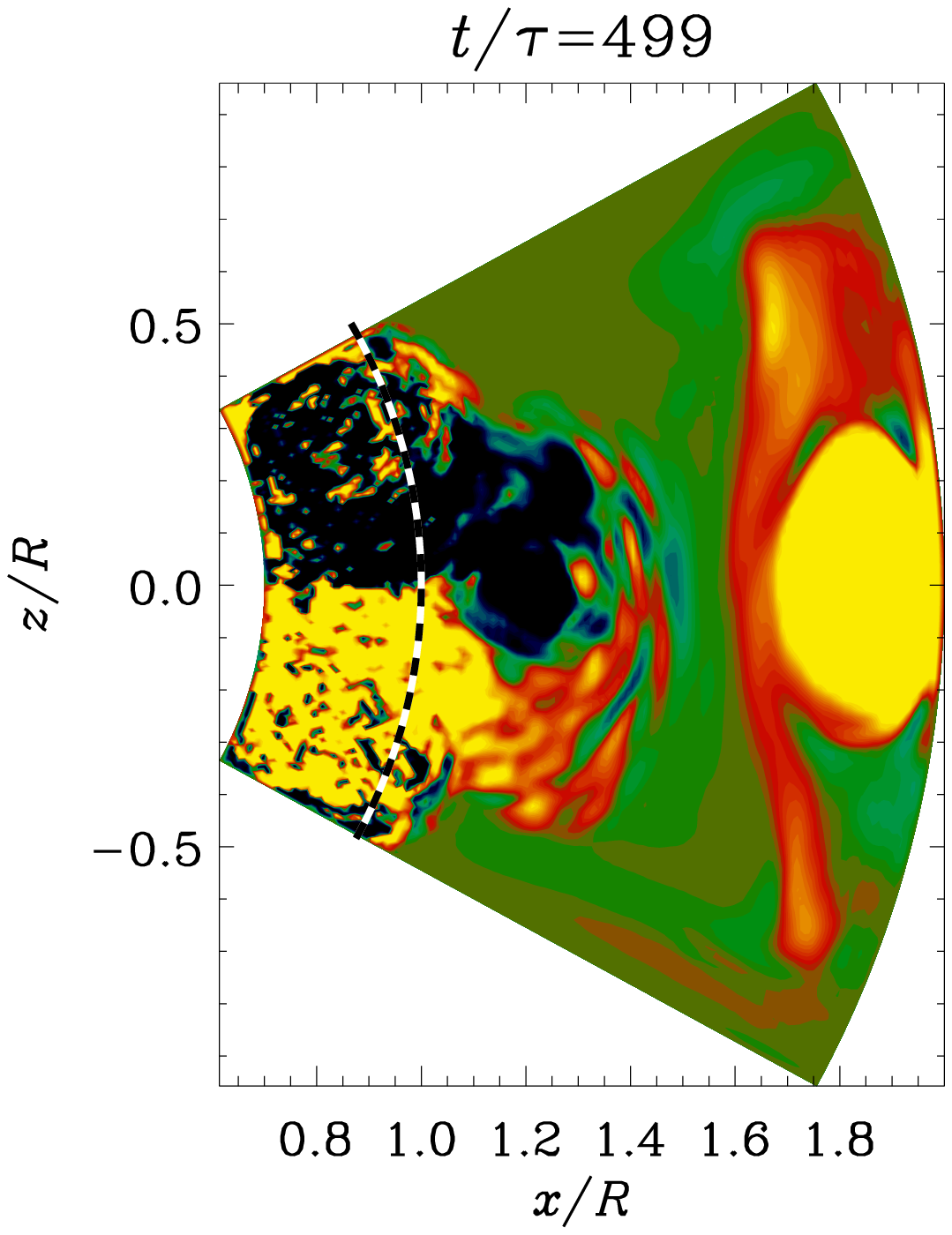}
\includegraphics[width=3.3cm]{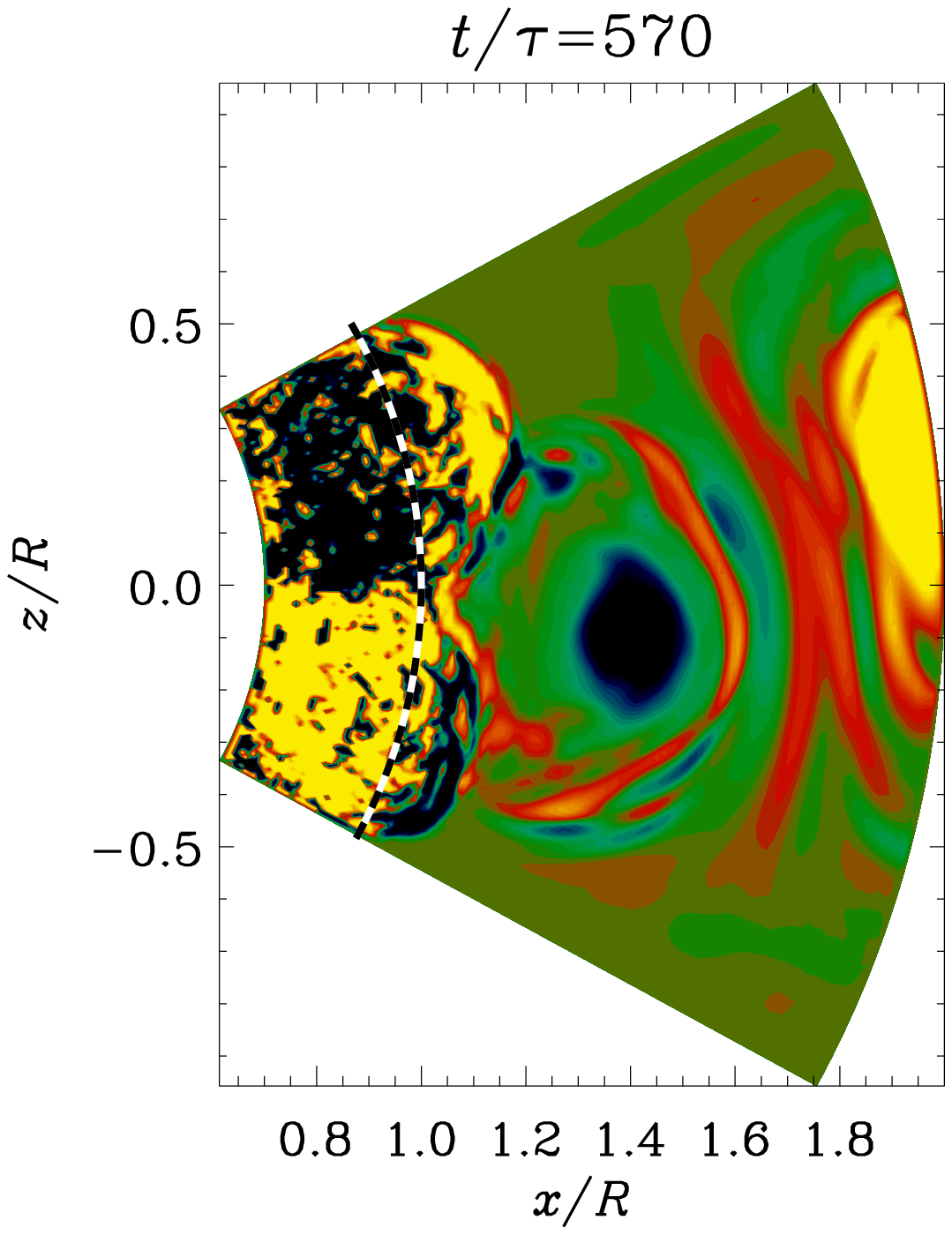}
\includegraphics[width=3.3cm]{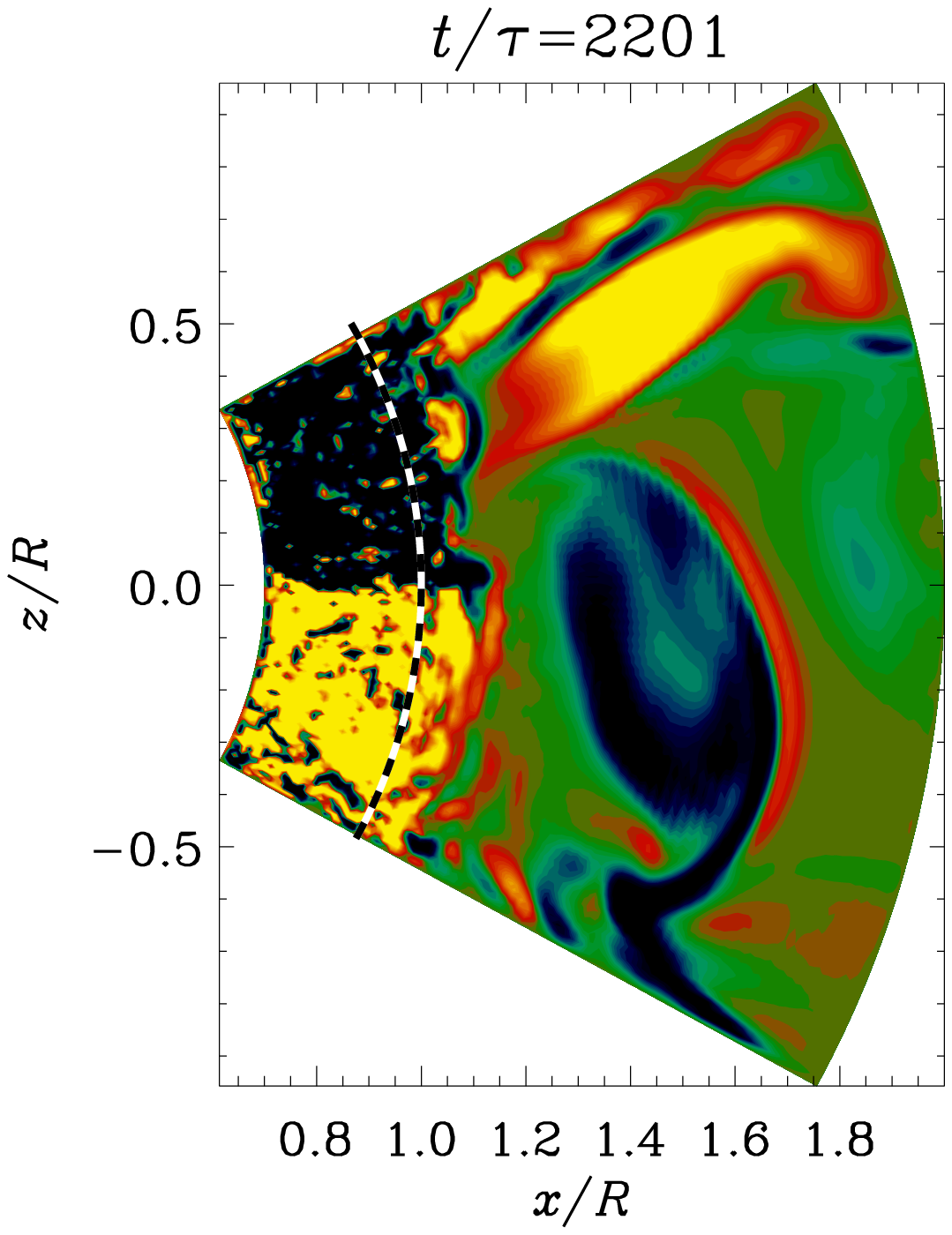}
\includegraphics[width=3.3cm]{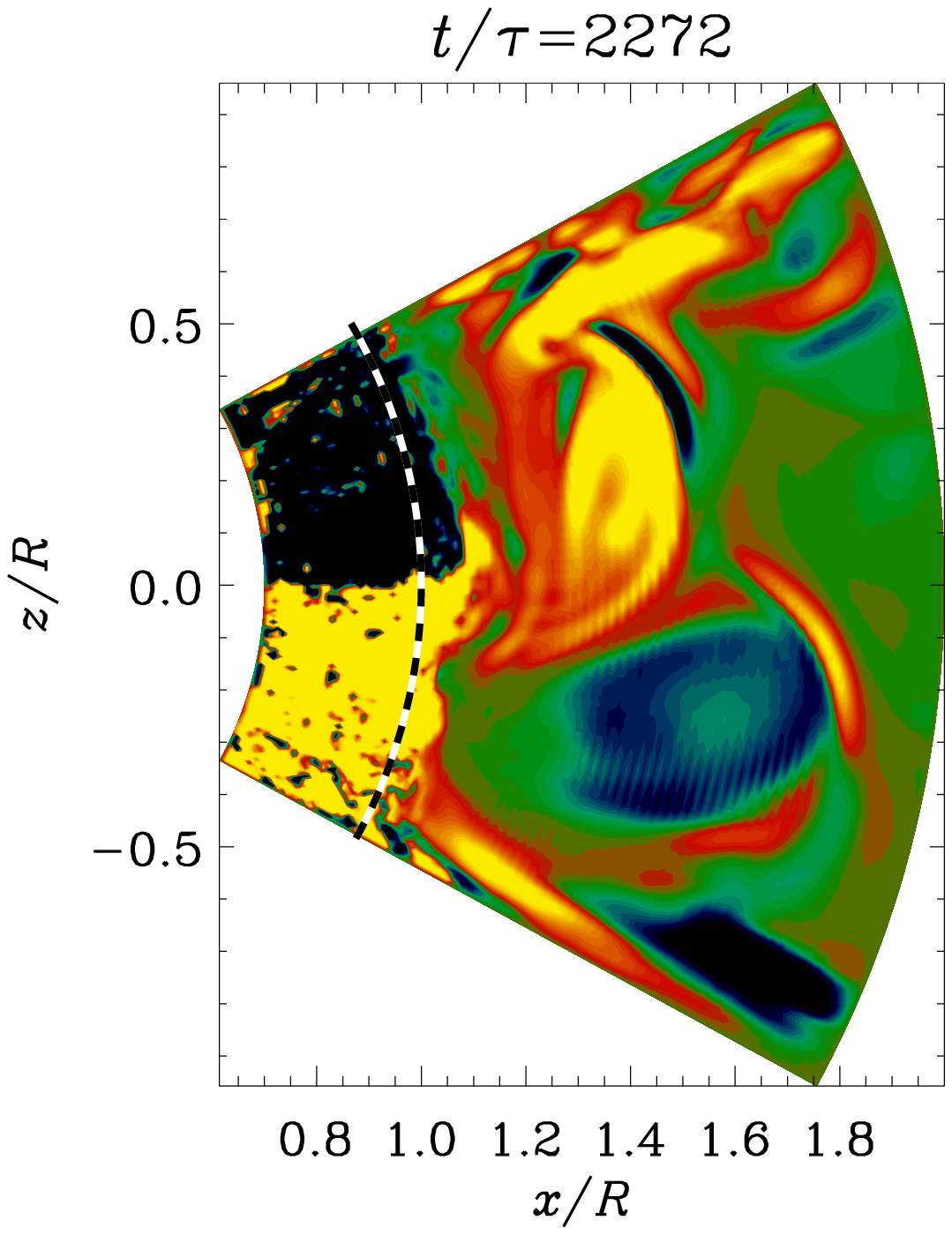}
\includegraphics[width=3.3cm]{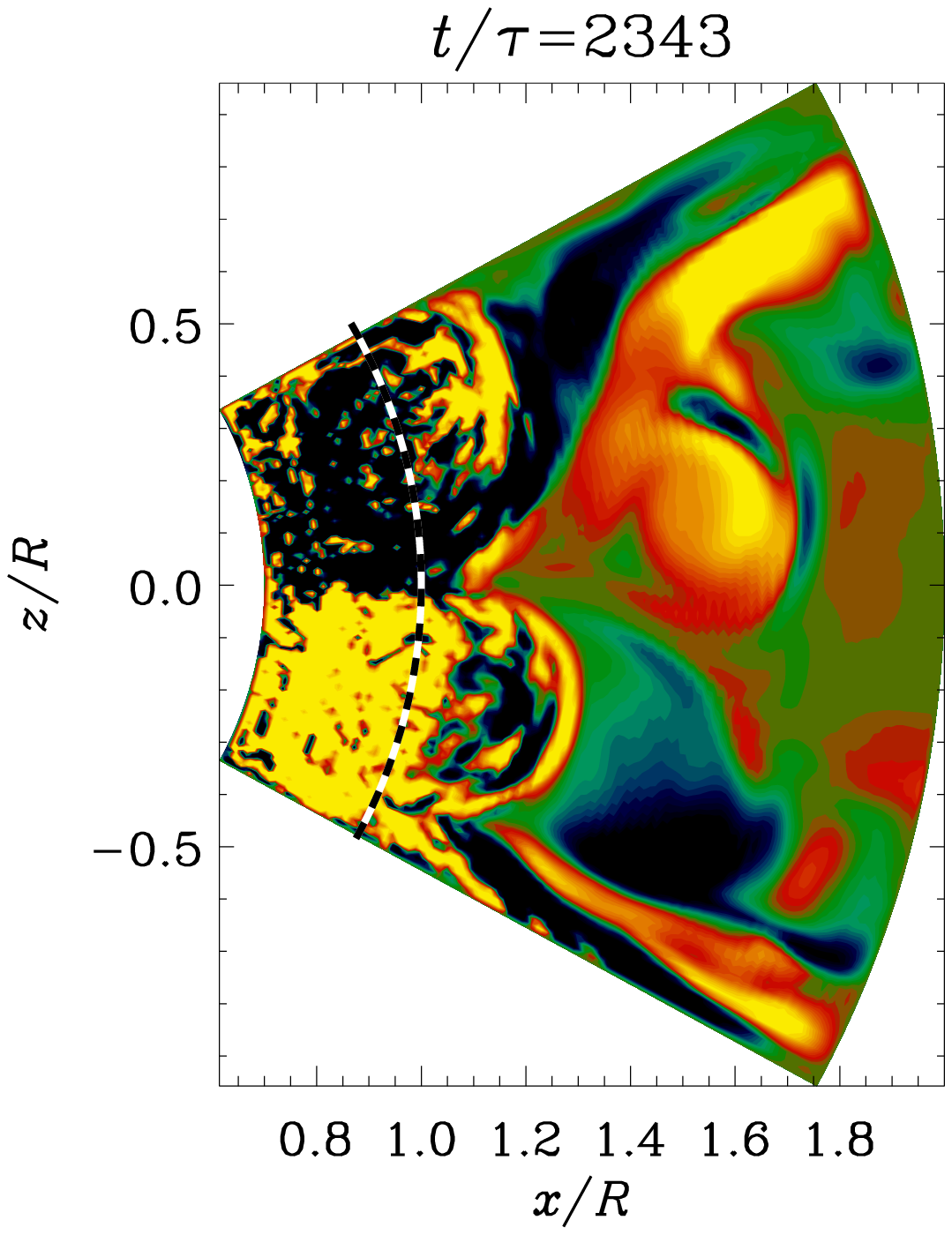}
\includegraphics[width=3.3cm]{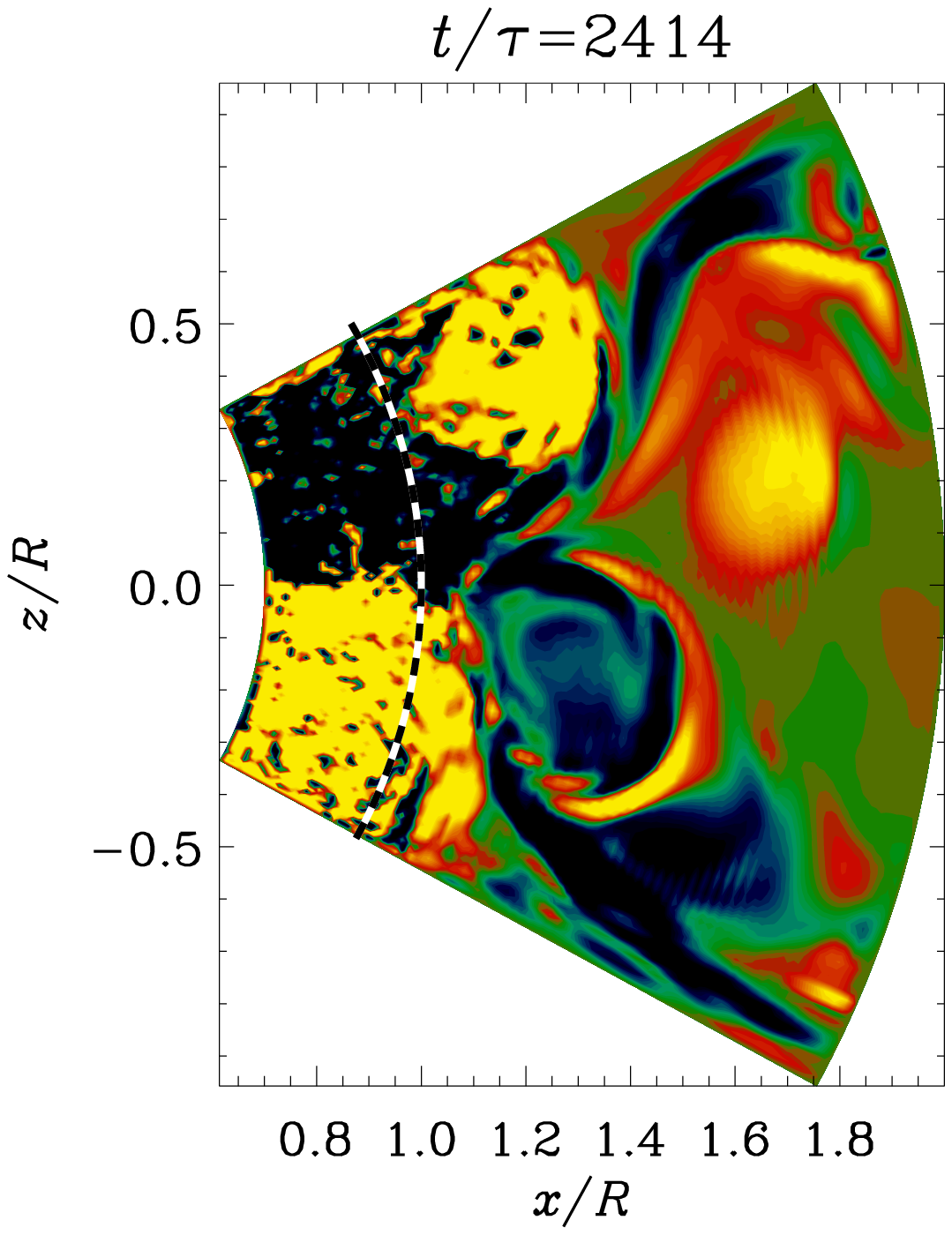}
\includegraphics[width=3.3cm]{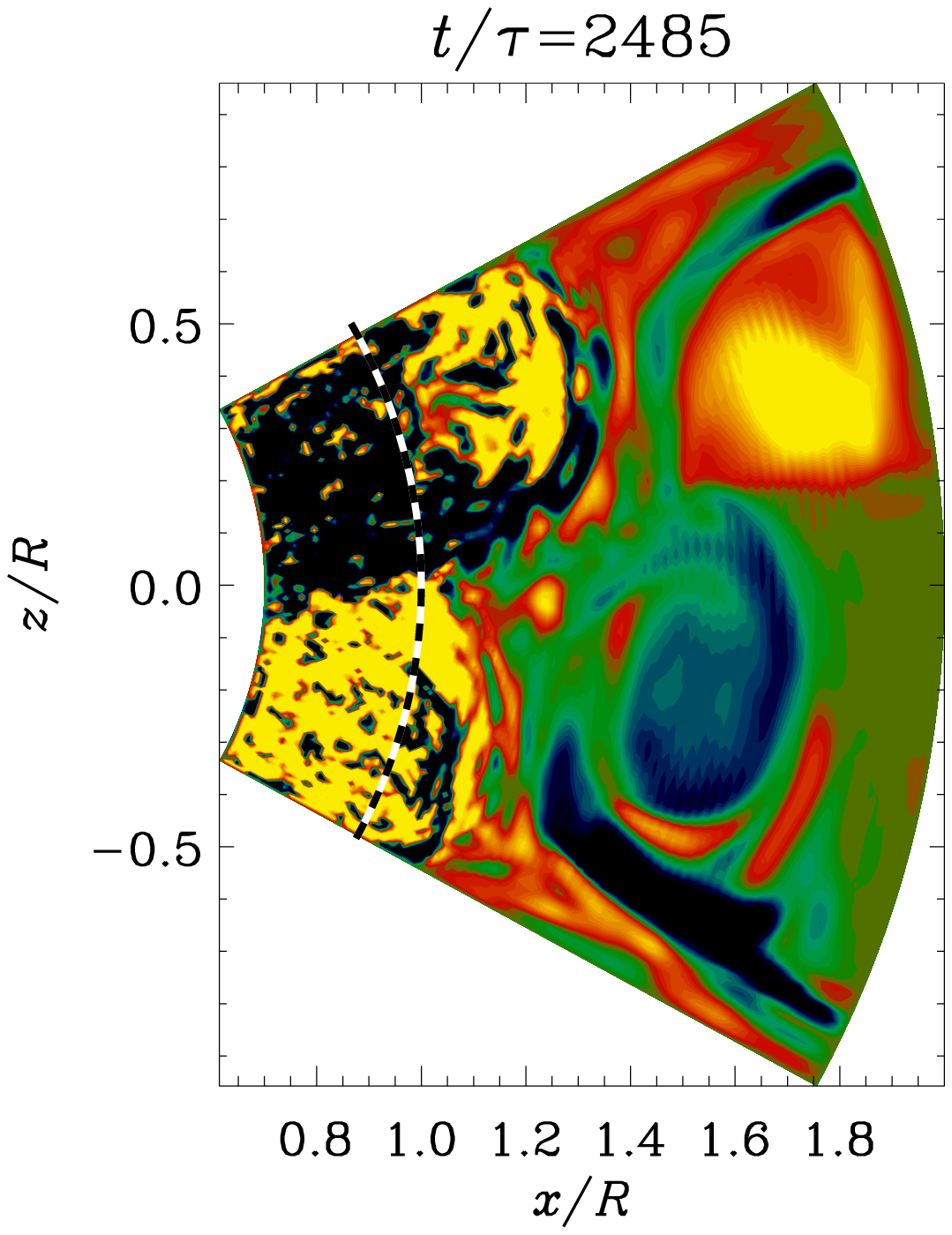}
\end{center}\caption[]{
Time series of coronal ejections in spherical coordinates.
The normalized current helicity,
$\overline{\JJ\cdot\BB}/\bra{\overline{\BB^2}}_t$, is shown in a
color-scale representation for different times; dark/blue stands for
negative and light/yellow for positive values, as in \Fig{pjbm_cont_TT_ct5}.
The dashed lines show the location of the surface at $r^2=x^2+z^2=R^2$.
}\label{jb_sph}
\end{figure*}

\begin{figure}[t!]
\begin{center}
\includegraphics[width=\columnwidth]{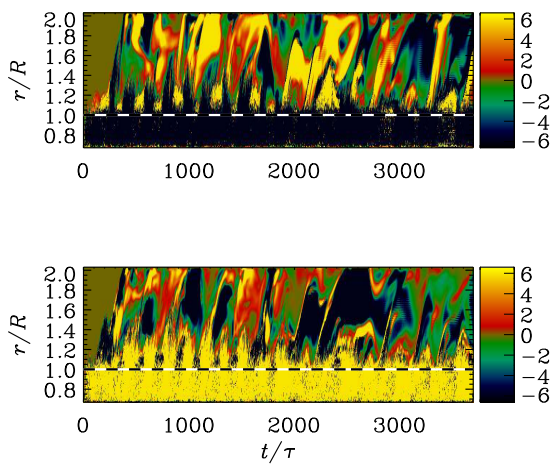}
\end{center}\caption[]{
Dependence of the dimensionless ratio
$\overline{\JJ\cdot\BB} / \bra{\overline{\BB^2}}_t$
on time and radius.
The top panel shows a narrow band in $\theta$ in the northern
hemisphere and the bottom one a narrow band in the southern
hemisphere.
Dark/blue stands for negative and light/yellow for positive values.
The
dashed
horizontal lines show the location of the surface at $r=R$.
}\label{pjbm3}
\end{figure}

While similar ejection events are also seen in spherical geometry,
a more surprising property is a reversal of the current helicity
in the radial direction \citep{WBM11}.
In \Fig{jb_sph} we show visualizations of $\JJ\cdot\BB$ in meridional
planes.
The Figure shows two examples of coronal ejections which we found to be
ejected from the dynamo region into the atmosphere.
 At $t/\tau=358$ we can identify a shape that is similar to that of the
three-part structure of coronal mass ejections, observed on the Sun \citep{Low96}.
It consists of an arch of one sign of current helicity in front of a bulk of
opposite sign and a cavity in between.
As show in \Fig{jb_sph}, the ejection leaves the domain on the radial
boundary and a new ejection of opposite sign occurs.
We also see that in the dynamo region the current helicity is negative
in the northern hemisphere and positive in the southern.
This seems to be basically true also in the immediate proximity above
the surface, but there is now an increasing tendency for the occurrence
of magnetic helicity of opposite sign ahead of the ejecta.
This seems to be associated with a redistribution of twist in the
swept-up material.
The current helicity is by far not always of the same sign, but both signs
occur and there is only a slight preference of one sign over the other.
This is seen more clearly in \Fig{pjbm3}, where we show a time series of
$\JJ\cdot\BB$ versus radial position $r/\Rsun$ and time $t/\tau$.

\begin{figure}[t!]
\begin{center}
\includegraphics[width=\columnwidth]{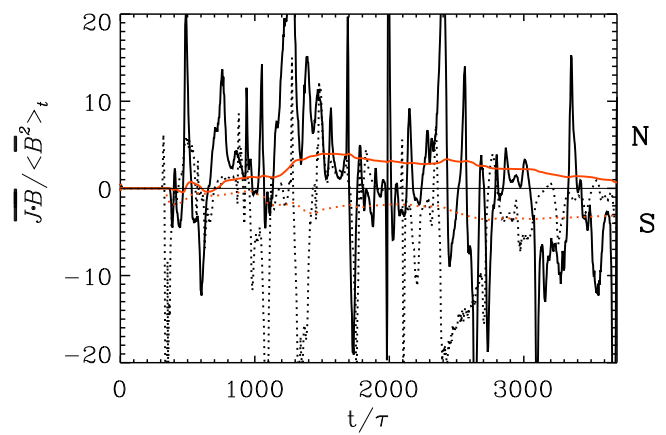}
\end{center}\caption[]{
Dependence of the dimensionless ratio
$\overline{\JJ\cdot\BB} / \bra{\overline{\BB^2}}_t$
on time at radius $r=1.7\,R$ and $28^{\circ}$ latitude.
The solid line stands for the northern hemisphere and the dotted for
the southern hemisphere.
The red lines represent the cumulative mean for each hemisphere.
}\label{pjbmP}
\end{figure}

\begin{figure}[t!]
\begin{center}
\includegraphics[width=5cm]{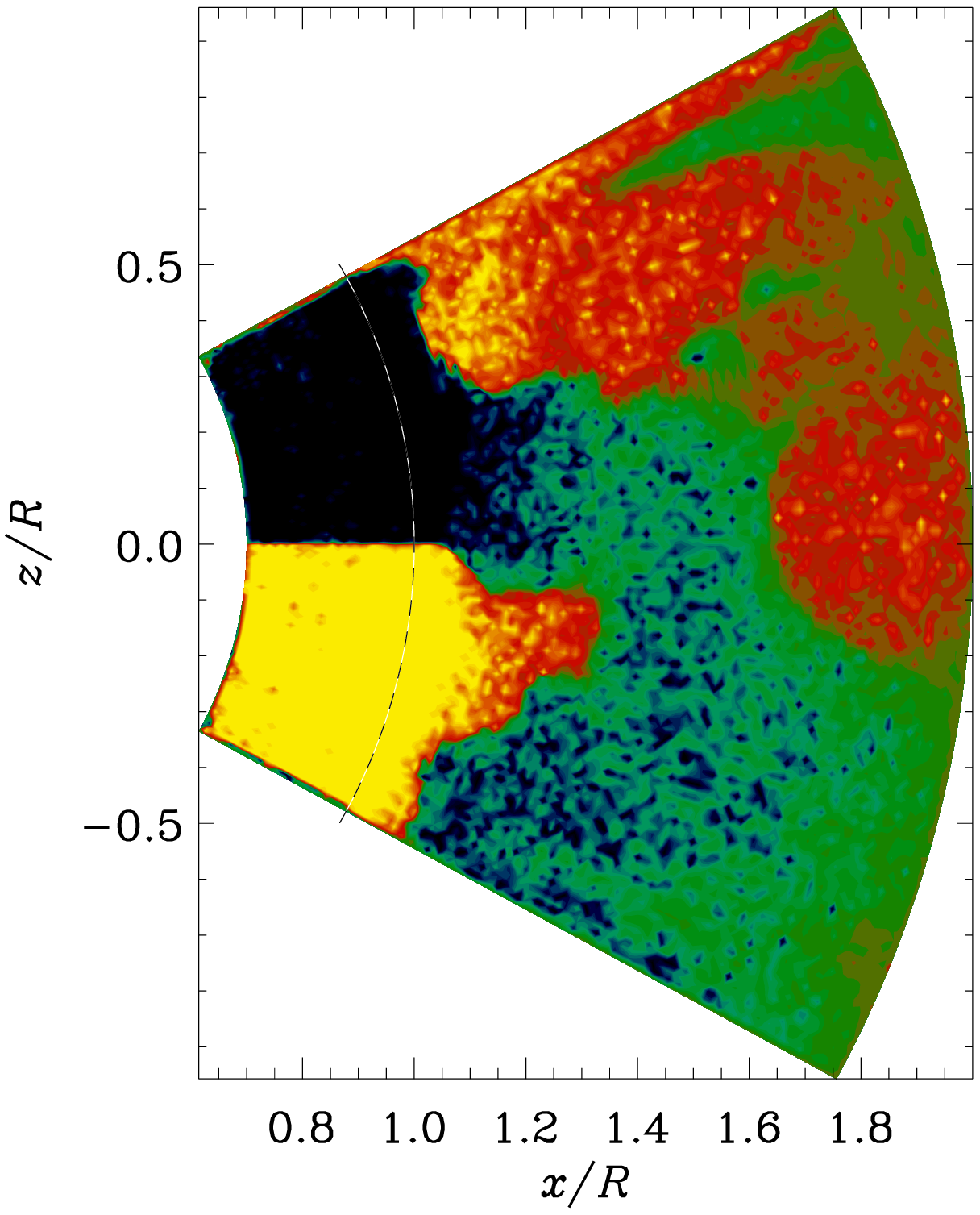}
\end{center}\caption[]{
Current helicity averaged over time.
Dark/blue corresponds to negative values, while the light/yellow corresponds
to positive value, as in \Fig{pjbm3}.
The dashed line show the location of the surface at $r^2=x^2+z^2=R^2$.
}\label{jbt}
\end{figure}

A time series of the normalized current helicity,
$\overline{\JJ\cdot\BB} / \bra{\overline{\BB^2}}_t$,
evaluated at radius $r=1.7\,R$ and $28^{\circ}$ latitude,
is shown in \Fig{pjbmP}, where we also show their running means.
It is now quite clear that on average the sign of current helicity
has changed relative to what it was in the dynamo region.
This is seen explicitly in a time-averaged plot of current helicity
in the meridional plane (\Fig{jbt}).

\begin{figure}[t!]
\begin{center}
\includegraphics[width=\columnwidth]{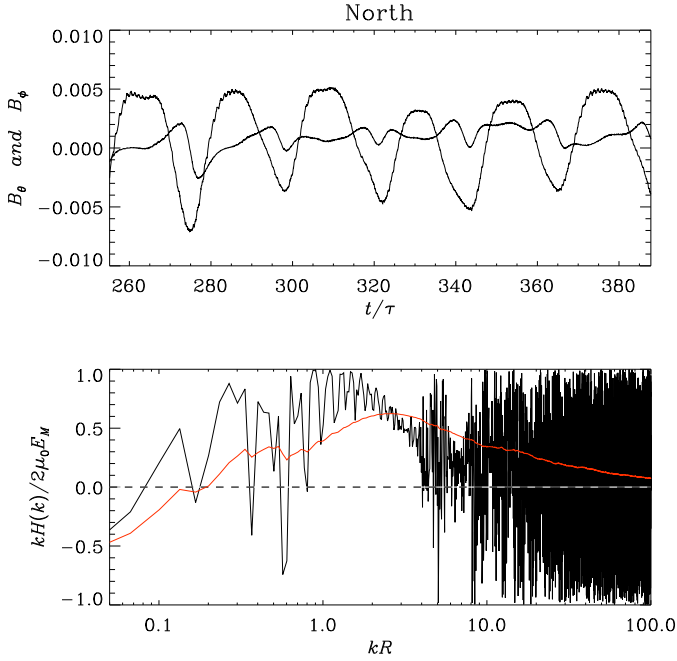}
\end{center}\caption[]{
Helicity in the northern outer atmosphere.
The values are written out at the point, $r=1.5\,R$,
$90^\circ-\theta=17^\circ$, and $\phi=9^\circ$.
{\it Top panel:} Phase relation between the toroidal $B_{\phi}$ and
poloidal $B_{\theta}$ field, plotted over time $t/\tau$.
{\it Bottom panel:} Helicity $H(k)$ is plotted over normalized wave number
$kR$.
The helicity
is calculated with the Taylor hypothesis using the
Fourier transformation of the poloidal and toroidal field.
Adapted from \cite{WBM11}.
}
\label{north}
\end{figure}

\begin{figure}[t!]\begin{center}
\includegraphics[width=\columnwidth]{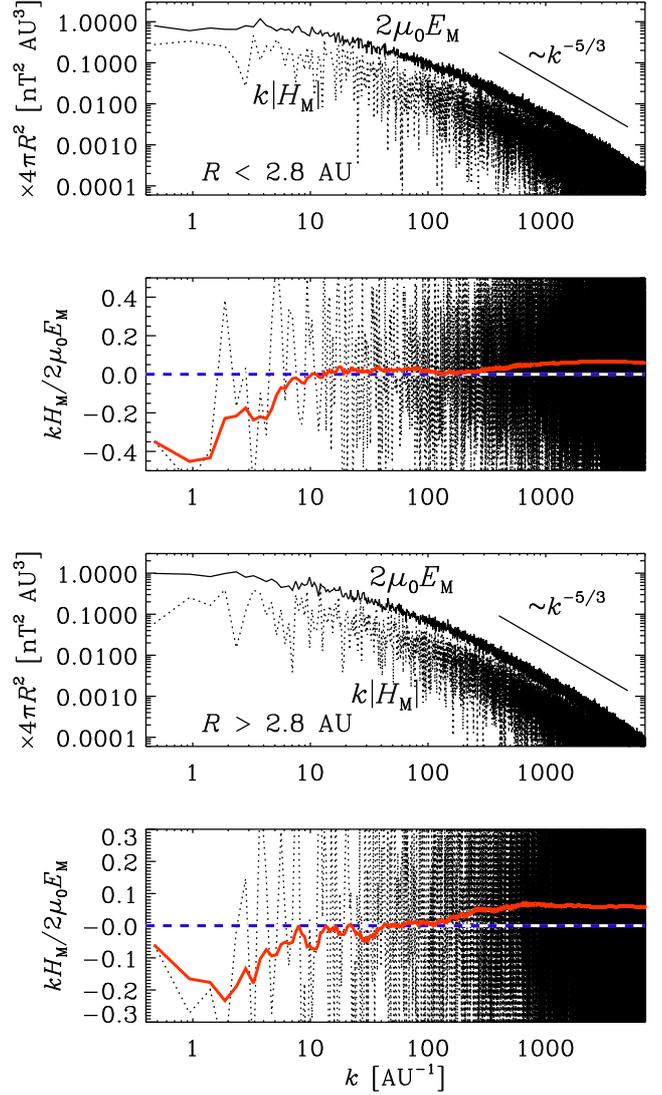}
\end{center}\caption[]{
Magnetic energy and helicity spectra, $2\mu_0 E_{\rm M}(k)$
and $k|H_{\rm M}(k)|$, respectively, for two separate distance intervals
(first and third panels).
Furthermore, both spectra are scaled by $4\pi R^2$ before averaging
within each distance interval above and below $2.8\AU$, respectively.
The relative magnetic helicity, $kH_{\rm M}(k)/2\mu_0 E_{\rm M}(k)$,
is plotted separately (second and fourth panels) together with its
cumulative average starting from the low wavenumber end.
The zero line is shown in dashed.
Adapted from \cite{BSBG11}.
}\label{rmeanhel4b_EH}\end{figure}

This reversal is significant because similar behavior has also been seen
in recent measurements of the magnetic helicity spectrum in the solar wind
\citep{BSBG11}, but before showing the evidence for this, let us first
discuss in more detail a similar diagnostics for the simulations.
In \Fig{north} we show the time variation of $B_\theta$ and $B_\phi$
and the magnetic helicity spectrum obtained from these time series.
Unlike the solar wind, where a time series can be used to mimic a scan
in distance space (the so-called Taylor hypothesis),
this argument fails in the present case, because no
wind is produced in the present model.
There is only the pattern speed associated with the CMEs.
Whether this is enough to motivate the use of the Taylor hypothesis
is rather unclear.
Nevertheless, there is a remarkable similarity with similar helicity
spectra obtained for the solar wind; see \Fig{rmeanhel4b_EH}.
In both cases, magnetic helicity has been obtained under the
assumption of local isotropy of the turbulence.
This means that one computes the one-dimensional
magnetic energy $E_{\rm M}^{\rm1D}(k_R)$ and
magnetic helicity spectra $H_{\rm M}^{\rm1D}(k_R)$ simply as
\EQ
E_{\rm M}(k_R)=|\hat{\BB}|^2/\mu_0,\quad
H_{\rm M}^{\rm1D}(k_R)=4\,{\rm Im}(\hat{B}_\theta\hat{B}_\phi^\star)/k_R,
\label{HelEqn}
\EN
where $k_R$ is the component of the wave vector in the radial direction.
Here, a hat denotes Fourier transformation and an asterisk complex conjugation.
These are the equations used by \cite{MGS82} who applied such an
analysis to data from {\it Voyager 2}.
Since {\it Voyager 2} flew close to the ecliptic, the magnetic helicity
is dominated by fluctuations.
This is why \cite{BSBG11} applied this analysis to {\it Ulysses} data,
where a net magnetic helicity was seen for the first time.
An important advantage of {\it Ulysses} over {\it Voyager 1} and {\it 2} is
the high angle with the ecliptic.
So, only with {\it Ulysses} we can measure the magnetic helicity
far away from the ecliptic in both hemispheres of the heliosphere.

\begin{table}[b!]\caption{
Distribution of magnetic helicity at large and small scales
both in the dynamo region close to or below
the surface and the corona region, solar wind or the exterior of the dynamo.
}\vspace{12pt}\centerline{\begin{tabular}{lcc}
Magnetic helicity & Large scales & small scales \\
\hline
Dynamo region, interior & $+$ & $-$ \\
Corona region, solar wind, exterior    & $-$ & $+$
\label{Tab}\end{tabular}}\end{table}

What we see both in the simulations and in the solar wind is that
there is magnetic helicity of opposite sign at large and small scales.
However, it is exactly the other way around than what is found in
the corona region; see \Tab{Tab}.
To understand the reason for this, we need to consider the equations
for the production of magnetic helicity at large and small length scales.
As will be argued in \Sec{ConnectionWith}, the mechanism that sustains
negative small-scale helicity in the north is turned off in the solar
wind, and there is just the effect of turbulent magnetic diffusion which
contributes with opposite sign.
By contrast, inside the dynamo region, turbulent diffusion is subdominant,
because otherwise no large-scale
magnetic field would be generated.
However, in the wind we do not expect the
dynamo to be exited, so here diffusion dominates.

\section{Connection with earlier dynamo models}
\label{ConnectionWith}

The purpose of this section is to make contact with dynamo theory
and to understand more quantitatively why the magnetic helicity
reverses sign with radius.
In essence, we argue that the profile of magnetic helicity density must
have a positive radial gradient to maintain a negative diffusive magnetic
helicity flux and that this is the reason for the magnetic helicity to
change from a negative sign to a positive one at some radius in the
northern hemisphere.

We begin by discussing the magnetic helicity equation.
The magnetic helicity density is $h=\AAA\cdot\BB$, where
$\BB=\nab\times\AAA$ is the magnetic field expressed in terms
of the magnetic vector potential $\AAA$ which, in the Weyl gauge,
satisfies $\partial\AAA/\partial t=\UU\times\BB-\eta\mu_0\JJ$.
The evolution equation for $\AAA\cdot\BB$ is then
\EQ 
{\partial\over\partial t}\AAA\cdot\BB=
-2\eta\mu_0\JJ\cdot\BB-\nab\cdot\FF,
\label{dAm}
\EN
where $\FF$ is the magnetic helicity flux.
Next, we define large-scale fields as averaged quantities, denoted by
an overbar, and small-scale fields as the residual,
denoted by lower case characters, so the magnetic
field can be split into two contributions via $\BB=\meanBB+\bb$.
Likewise, $\AAA=\meanAA+\aaaa$, $\JJ=\meanJJ+\jj$, and $\UU=\meanUU+\uu$.
The evolution equation for the mean magnetic helicity
density $\meanh=\overline{\AAA\cdot\BB}$ is given by
\EQ 
{\partial\over\partial t}\overline{\AAA\cdot\BB}=
-2\eta\mu_0\overline{\JJ\cdot\BB}-\nab\cdot\meanFF.
\label{dABm}
\EN
To determine the magnetic helicity density of the mean field,
$\meanhm=\meanAA\cdot\meanBB$,
we use the averaged induction equation in the Weyl gauge,
$\partial\meanAA/\partial t=\meanUU\times\meanBB+\overline{\uu\times\bb}
-\eta\mu_0\meanJJ$, so that
\EQ
{\partial\over\partial t}\meanAA\cdot\meanBB=
2\,\overline{\uu\times\bb}\cdot\meanBB
-2\eta\mu_0\meanJJ\cdot\meanBB-\nab\cdot\meanFFm.
\label{dAmBm}
\EN
The magnetic helicity equation for the fluctuating field,
$\meanhf=\meanh-\meanhm=\overline{\aaaa\cdot\bb}$, takes then the form
\EQ 
{\partial\over\partial t}\overline{\aaaa\cdot\bb}=
-2\,\overline{\uu\times\bb}\cdot\meanBB
-2\eta\mu_0\overline{\jj\cdot\bb}-\nab\cdot\meanFFf,
\label{dabm}
\EN
so that the sum of \Eqs{dAmBm}{dabm} gives \Eq{dABm}.
Here, the total magnetic helicity flux consists of contribution from mean
and fluctuating fields, denoted by subscripts m and f, respectively, i.e.,
$\meanFFm+\meanFFf=\meanFF$.
Note that, even in the limit $\eta\to0$ and in the absence of fluxes,
magnetic helicity at large and small scales is not conserved individually,
but there can be an exchange of magnetic helicity between scales.

A note regarding the gauge-dependence is here in order.
Obviously, \Eq{dabm} depends on the gauge choice for $\aaaa$.
However, if we are in a steady state, and if $\meanhf$ also happens
to be steady (which is not automatically guaranteed), then we have
\EQ
\nab\cdot\meanFFf=-2\,\overline{\uu\times\bb}\cdot\meanBB
-2\eta\mu_0\overline{\jj\cdot\bb},
\EN
and since the right-hand side of this equation is manifestly
gauge-invariant, $\nab\cdot\meanFFf$ must also be gauge-invariant.
This property was used in earlier work of \cite{Mitra}, \cite{HB10} and
\cite{WBM11} to determine
the scaling of $\meanFFf$ with $\nab\meanhf$ and thus the turbulent
diffusion coefficient $\kappa_h$.
In addition, if there is sufficient scale separation between large
and small scales, which is typically the case in the nonlinear regime
at the end of the inverse cascade process \citep{B01}, then
$\meanhf$ can be expressed as a density of linkages, which is itself
manifestly gauge-independent \citep{SB06}.
This property then also applies to the flux $\meanFFf$.

The correlation $\overline{\uu\times\bb}$ is known to have two contributions,
one proportional to $\meanBB$ with a pseudo-tensor in front of it
(the $\alpha$ effect, responsible for large-scale field generation),
and one proportional to $\meanJJ$ with a coefficient $\etat$ in front of it
that corresponds to turbulent diffusion, i.e.,
\EQ
\overline{\uu\times\bb}=\alpha\meanBB-\etat\mu_0\meanJJ,
\EN
where we have again assumed isotropy.
The reason why the mean magnetic helicity density of the small-scale field is
negative in the north is because $\alpha>0$ in the north \citep[e.g.][]{KR80},
producing therefore negative magnetic helicity at a rate
$-2\alpha\meanBB^2<0$ for small-scale fields and
$+2\alpha\meanBB^2>0$ for large-scale fields, so that their sum
vanishes.
There is also turbulent magnetic diffusion which reduces this effect,
because $\etat>0$ and $\meanJJ\cdot\meanBB>0$ in the north.
In the solar wind no new magnetic field is generated,
so turbulent magnetic diffusion could now dominate and might thus
explain a reversal of magnetic helicity density \citep{BSBG11}.
 
\begin{figure}[t!]\begin{center}
\includegraphics[width=\columnwidth]{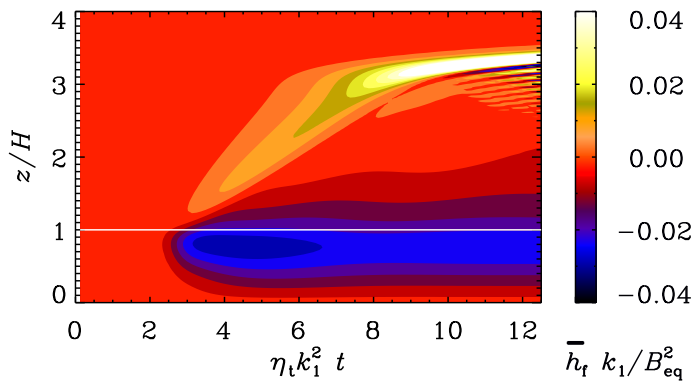}
\end{center}\caption[]{
$zt$ diagrams of $\meanhf$ for an $\alpha^2$ dynamo with a wind
which stops all of a sudden at $z/H=3$.
The white horizontal line marks the location $z=H$.
Light/yellow shades indicate positive values and dark/blue shades
indicate negative values.
}\label{pbutter_hel_alpprof3_Uprof4_shock}\end{figure}

Support for a reversal of the sign of magnetic helicity was first seen
in dynamo simulations with magnetic helicity flux in the exterior.
In \Fig{pbutter_hel_alpprof3_Uprof4_shock} we show a representation
of magnetic helicity density of small-scale fields
$\meanhf=\overline{\aaaa\cdot\bb}$ versus $z$ and $t$ for a model
similar
to that of \cite{BCC09}, but where the magnetic helicity flux is
caused by a wind that is then running into a shock\footnote{Note that
the shock at $z/H=3$ becomes eventually under-resolved and
the simulation has to be terminated.
This is what causes the wiggles in the proximity of the shock.} where
the flux is artificially suppressed at height $z=3H$.
This figure shows that there is a clear segregation of negative
and positive small-scale magnetic helicity in the dynamo regime
and the exterior, respectively.

\begin{figure}[t!]\begin{center}
\includegraphics[width=\columnwidth]{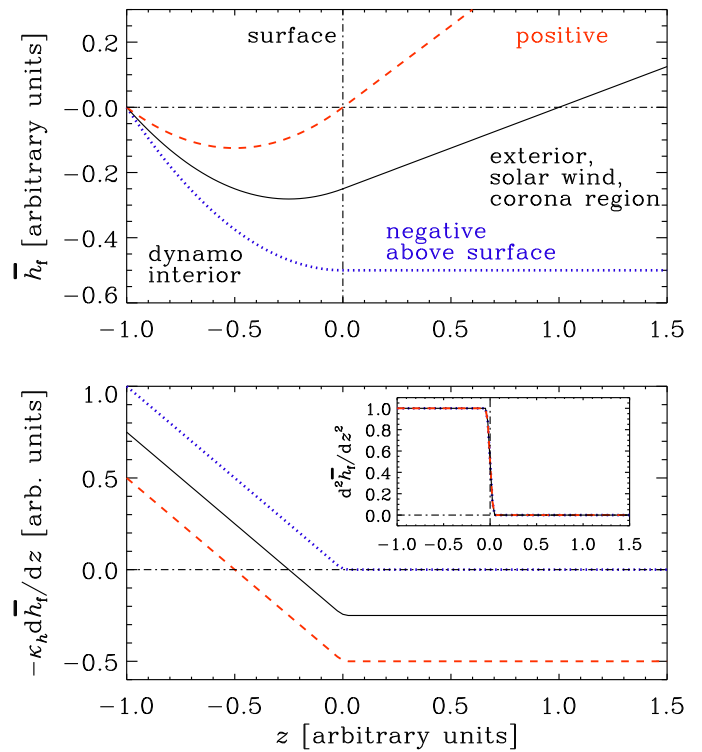}
\end{center}\caption[]{
Sketch showing possible solutions $\meanhf(z)$ (upper panel) to
\Eq{diffeqn} with $S=\const=-1$ in $z<0$ and $S=0$ in $z>0$.
The red (dashed) and black (solid) lines show solutions for which the
magnetic helicity flux ($-\kappa_h\dd\meanhf/\dd z$, see lower panel)
is negative in the exterior.
This corresponds to the case observed in the Sun.
The blue (dotted) line shows the case, where the magnetic helicity
flux is zero above the surface and therefore do not reverse the sign of
$\meanhf(z)$ in the exterior.
}\label{psketch}\end{figure}

In our description above we have suggested that the magnetic helicity
production balances the $\partial\overline{\aaaa\cdot\bb}/\dd t$ term,
but this cannot be true in the steady state.
Instead, it must be the divergence of the magnetic helicity flux,
$\nab\cdot\meanFFf$.
Let us assume that $\meanFFf$ can be approximated by a Fickian
diffusion law, i.e., $\meanFFf=-\kappa_h\nab\meanhf$.
Simulations have suggested that $\kappa_h/\etat$ is around 0.3
\citep{Mitra,HB10,WBM11}.
Thus, balancing now the source
$S(z)\equiv-2\alpha\meanBB^2+2\etat\mu_0\overline{\JJ \cdot \BB}$
against the divergence of the flux of magnetic helicity at small scales,
we have, in a one-dimensional model (neglecting the molecular
diffusion term, $2\eta\mu_0\overline{\jj\cdot\bb}$):
\EQ
S(z)=-\kappa_h{\dd^2\meanhf\over\dd z^2}.
\label{diffeqn}
\EN
Taking as an example a source where $S(z)/\kappa_h=-1$ in the
dynamo interior ($z<0$), and $S(z)=0$ in $z>0$, we have a family
of solutions of \Eq{diffeqn} that only differ in an undetermined
integration constant corresponding to a constant offset in the flux;
see the second panel of \Fig{psketch}.
The solutions for which the magnetic helicity flux,
$-\kappa_h\dd\meanhf/\dd z$, is negative in the exterior
are those for which $\meanhf$ reaches an extremum below the surface.
This seems to be what happens both in simulations and in the solar wind.
We can thus conclude that the reason for a sign change is not a dominance
of turbulent diffusion in the solar wind, but just the possibility of the
magnetic helicity density reaching an extremum {\it below} the surface
(dashed red and solid black lines in \Fig{psketch}), not at the
surface (dotted blue lines).

\section{Conclusions}

As we have seen in the present paper, magnetic twist (or helicity) plays
an important role for the solar dynamo \citep{B01,BB02} and for producing
eruptions of the form of CMEs \citep{Low96,Low01}.
The recent work of \cite{WB10} and \cite{WBM11} tries to combine both
aspects into one.
Although the models are still rather unrealistic in many respects, they
have already now led to useful insights into the interplay between dynamo
models and solar wind turbulence.
In particular, they have allowed us to understand the properties
of magnetic helicity fluxes.
We have confirmed that the hemispheric sign rule of magnetic helicity
does not extend unchanged into the interplanetary space, but we have
now shown that it must flip sign somewhere above the solar surface.
On the other hand, \cite{BS98} found that the magnetic clouds follow Hale's
polarity and that the sign of the magnetic helicity is the same as in the
interior. However, this result is not based on rigorous statistics.

Future work in this direction should include more realistic modeling
of the solar convection zone.
Preliminary work in this direction is already underway \citep{WKMB12}.
Furthermore, it will be necessary to allow for the development of
a proper solar wind from the dynamo region.
One of the difficulties here is that, if the critical point is assumed
to be too close to the solar surface, which would be computationally
convenient because it would allow us to use a smaller domain,
the mass loss rate would be rather high and could destroy the dynamo.
In future work we will be trying to strike an appropriate compromise that
will allow us to study the qualitatively new effects emerging from this.

\begin{acknowledgements}
We acknowledge the allocation of computing resources provided by the
Swedish National Allocations Committee at the Center for
Parallel Computers at the Royal Institute of Technology in
Stockholm, the National Supercomputer Centers in Link\"oping and the
High Performance Computing Center North in Ume\aa.
This work was supported in part by
the European Research Council under the AstroDyn Research
Project No.\ 227952 and the Swedish Research Council Grant No.\ 621-2007-4064,
and the National Science Foundation under Grant No.\ NSF PHY05-51164.
\end{acknowledgements}


\vfill\bigskip\noindent\tiny\begin{verbatim}

$Header: /var/cvs/brandenb/tex/joern/namur/paper.tex,v 1.62 2012-07-15 21:12:36 joern Exp $
\end{verbatim}

\begin{thebibliography}{}

\bibitem[Blackman \& Brandenburg(2002)]{BB02}
Blackman, E. G., \& Brandenburg, A.\yapj{2002}{579}{359}

\bibitem[Blackman \& Brandenburg(2003)]{BB03}
Blackman, E. G., \& Brandenburg, A.\yapjl{2003}{584}{L99}

\bibitem[Bothmer \& Schwenn(1998)]{BS98}
Bothmer, V. \& Schwenn, R. {1998, {Ann.\ Geophysicae}, {16}, 1}

\bibitem[Brandenburg(2001)]{B01}
Brandenburg, A.\yapj{2001}{550}{824}

\bibitem[Brandenburg et al.(2009)]{BCC09}
Brandenburg, A., Candelaresi, S., \& Chatterjee, P.\ymn{2009}{398}{1414}

\bibitem[Brandenburg et al.(2011)]{BSBG11}
Brandenburg, A., Subramanian, K., Balogh, A., \& Goldstein, M. L.\yapj{2011}{734}{9}

\bibitem[Hubbard \& Brandenburg(2010)]{HB10}
Hubbard, A., \& Brandenburg, A.\ygafd{2010}{104}{577}

\bibitem[K\"apyl\"a et al.(2010)]{KKBMT10}
K\"apyl\"a, P.J., Korpi, M. J., Brandenburg, A., Mitra, D. \& Tavakol, R. \yan{2010}
{331}{73}

\bibitem[K\"apyl\"a et al.(2012)]{KMB12}
K\"apyl\"a, P.J., Mantere, M.J. \& Brandenburg, A.\sapjl{2012},
{arXiv:1205.4719}

\bibitem[Krause and R\"adler(1980)]{KR80}
Krause, F., R\"adler, K.-H.\ybook{1980}
{Mean-field magneto\-hydro\-dy\-na\-mics and dynamo theory}
{Pergamon Press, Oxford}

\bibitem[Low(1996)]{Low96}
Low, B. C.\ysph{1996}{167}{217}

\bibitem[Low(2001)]{Low01}
Low, B. C.\yjgr{2001}{106}{25,141}

\bibitem[Matthaeus et al.(1982)]{MGS82}
Matthaeus, W. H., Goldstein, M. L., \& Smith, C.\yprl{1982}{48}{1256}

\bibitem[Mitra et al.(2010)]{Mitra}
Mitra, D., Candelaresi, S., Chatterjee, P., Tavakol, R., \& Brandenburg, A.\yan{2010}{331}{130}

\bibitem[Ortolani \& Schnack(1993)]{OS93}
Ortolani, S., \& Schnack, D. D.\ybook{1993}{Magnetohydrodynamics of plasma
relaxation}{World Scientific, Singapore}

\bibitem[Pouquet et al.(1976)]{PFL}
Pouquet, A., Frisch, U., \& L\'eorat, J.\yjfm{1976}{77}{321}

\bibitem[Subramanian \& Brandenburg(2006)]{SB06}
Subramanian, K., \& Brandenburg, A.\yapj{2006}{648}{L71}

\bibitem[Warnecke \& Brandenburg(2010)]{WB10}
Warnecke, J., \& Brandenburg, A.\yana{2010}{523}{A19}

\bibitem[Warnecke et al.(2011)]{WBM11}
Warnecke, J., Brandenburg, A., \& Mitra, D.\yana{2011}{534}{A11}

\bibitem[Warnecke et al.(2012)]{WKMB12}
Warnecke, J., K\"apyl\"a, P. J., Mantere, M. J., \& Brandenburg, A.\ysol{2012}
{in press}{arXiv:1112.0505}
\end{thebibliography}
\end{document}